\documentclass[reprint,amsmath,amssymb,aps]{revtex4-2}

\usepackage{graphicx}
\usepackage{dcolumn}
\usepackage{bm}
\usepackage{xcolor}
\usepackage{braket}
\usepackage[colorlinks=true, allcolors=blue]{hyperref}

\begin{document}

\title{Effects of Small-Chain Superexchange Dynamics on \\ Spin-Orbit Coupled Clock Spectroscopy}% Force line breaks with \\

\author{Mikhail Mamaev \\}
 \affiliation{Pritzker School of Molecular Engineering, University of Chicago, Chicago, Illinois, USA}

\author{Ana M. Rey}
 \affiliation{JILA, National Institute of Standards and Technology and the University of Colorado, Boulder, Colorado, USA
and Department of Physics, University of Colorado, Boulder, Colorado, USA}

\author{William R. Milner\\}
 \affiliation{Department of Physics, Massachusetts Institute of Technology, Cambridge, Massachusetts, USA}

\date{\today}

\begin{abstract}

Optical lattice clocks have set records in clock precision and accuracy.  Continuing to advance their performance, via probing as many atoms for the longest interrogation time affordable, requires experimentally and theoretically studying a many-body lattice system. Motivated by recent experimental results on a Fermi-degenerate three-dimensional optical lattice clock~\cite{milner_science}, we present a theoretical overview of Ramsey and Rabi spectroscopy in one-dimensional chains. At realistic experimental temperatures and confinement conditions, atoms are spatially localized into small chains of $\approx 1-5$ atoms. We show that in the presence of spin-orbit coupling induced by the clock laser, the spectroscopy observables are modified by superexchange interactions within each chain, and depend strongly on the length of the chain. The thermal distribution of chain lengths thus plays a key role in the spectroscopy measurements. Our results offer insight into  observable many-body effects in state-of-the-art lattice clocks and suggest new directions for optimizing clock performance.
\end{abstract}

\maketitle

%%%%%
\section{Introduction}
%%%%%

\indent Improving the precision of optical lattice clocks will open the door to increasingly stringent tests of fundamental physics~\cite{kolkowitz2016gravitational, sanner_lorentz, kennedy2020precision, katori_GR, rmp_sensing_2} and demand understanding novel clock systematics at the $10^{-19}$ level~\cite{chang2004controlling, hutson2023observation}. Limited by quantum projection noise, improving clock precision requires probing as many atoms as possible for the longest coherent interrogation times affordable. For the current state-of-the-art employing hundreds of thousands of atoms probed for tens of seconds~\cite{bothwell2022resolving, milner_science}, improvements demand experimentally studying and theoretically modeling a complex, interacting many-body physics system. These interactions can be harnessed to generate spin entanglement for improved clock precision~\cite{Robinson2023entanglement, eckner2023realizing}. 

Confining atoms in the ground band of a three-dimensional lattice is an attractive solution to optimize clock stability. The resulting system admits a description using the canonical Fermi-Hubbard model, and enables scalability of atom number without necessarily introducing detrimental errors from atomic interactions. Although $1.5 \times 10^{-17}/\sqrt{t}$ clock stability (for interrogation time $t$) has been achieved on this experimental platform, there are outstanding challenges that limit atomic coherence times at the $\sim 10$ second level. At shallow lattice depths, atomic coherence times are limited by motional dephasing when atoms tunnel site-to-site~\cite{lemonde2005optical, kolkowitz2017spin}.  As lattice depths are increased, optical lattice induced Raman scattering destroys the delicate superposition states required for clock spectroscopy~\cite{dorscher2018lattice, hutson2019engineering}. Thus to achieve the optimal coherence time, one must operate in a regime of intermediate lattice depth where atoms are spatially localized, but effects of tunneling are not completely negligible. 

In general, this condition can be fulfilled in the Mott insulating regime at half-filling, where repulsive contact interactions localize atoms when the interaction strength $U$ exceeds the tunneling coupling $t$~\cite{duan2003controlling, trotzky2008time}. In this regime, the atoms undergo effective spin interactions induced by the superexchange mechanism at a rate $V \propto 4 t^2 / U$. Due to spin-orbit coupling (SOC)~\cite{celi2014synthetic, barbarino2016synthetic} imparted by the clock laser~\cite{kolkowitz2017spin}, these spin interactions break the underlying SU(2) symmetry of the interactions and realize an anisotropic spin Hamiltonian. This anisotropy generates interaction-induced many-body dynamics. The effects of imprinted spin-spiral textures on spin transport and relaxation have been studied extensively in optical lattice experiments~\cite{aidelsburger2013realization, livi2016synthetic,tai2017microscopy, hild2014far, jepsen2021transverse, jepsen2022long} and Bose-Einstein condensates~\cite{liang2021coherence, an2017direct}. These dynamical properties can strongly modify Ramsey and Rabi clock spectroscopy protocols for optical lattice clocks, as the superexchange coupling can reach the Hz level or more. The anisotropic spin interactions can also lead to dephasing of the atomic  coherence which was experimentally observed in a 3D optical lattice clock~\cite{milner_science}. Both theoretical and experimental explorations into the dynamics of atoms in this regime are required to advance optical lattice clocks at high density.

In this work, we theoretically study Ramsey and Rabi spectroscopy protocols for atoms confined in one dimensional chains in the interaction dominated regime. For typical experimental temperatures and confinement conditions, 
atoms are localized into small chains of $\approx 1 - 5$ atoms, sub-divided by empty sites arising from the finite initial temperature of the system. We show that the observed dynamics can be modeled by contributions from decoupled spin chains of various lengths. Most importantly, we show that in the presence of SOC, these dynamics strongly depend on the number of atoms in the chain. We characterize the expected spectroscopy signal in these chains, and study the dependence of the spectroscopy signal on temperature and interaction strength.

The paper is organized as follows. In Section~\ref{opticallattice_sec}, we introduce the optical lattice setup and we detail the system's theoretical description by a Fermi-Hubbard model. In Section~\ref{superex_sec} we provide a low-energy superexchange spin-1/2 description of the system. In Sections~\ref{ramsey_sec} and~\ref{rabi_sec} we study Ramsey and Rabi spectroscopy protocols in the presence of superexchange interactions. In Section~\ref{sec_Holes} we compare the physics of the full Fermi-Hubbard model to the superexchange model, as well as a $t-J$ model that better captures the effects of holes.

%%%%%
\section{3D optical lattice setup}
\label{opticallattice_sec}
%%%%%

\begin{figure*}
\center
\includegraphics[width=0.85\textwidth]{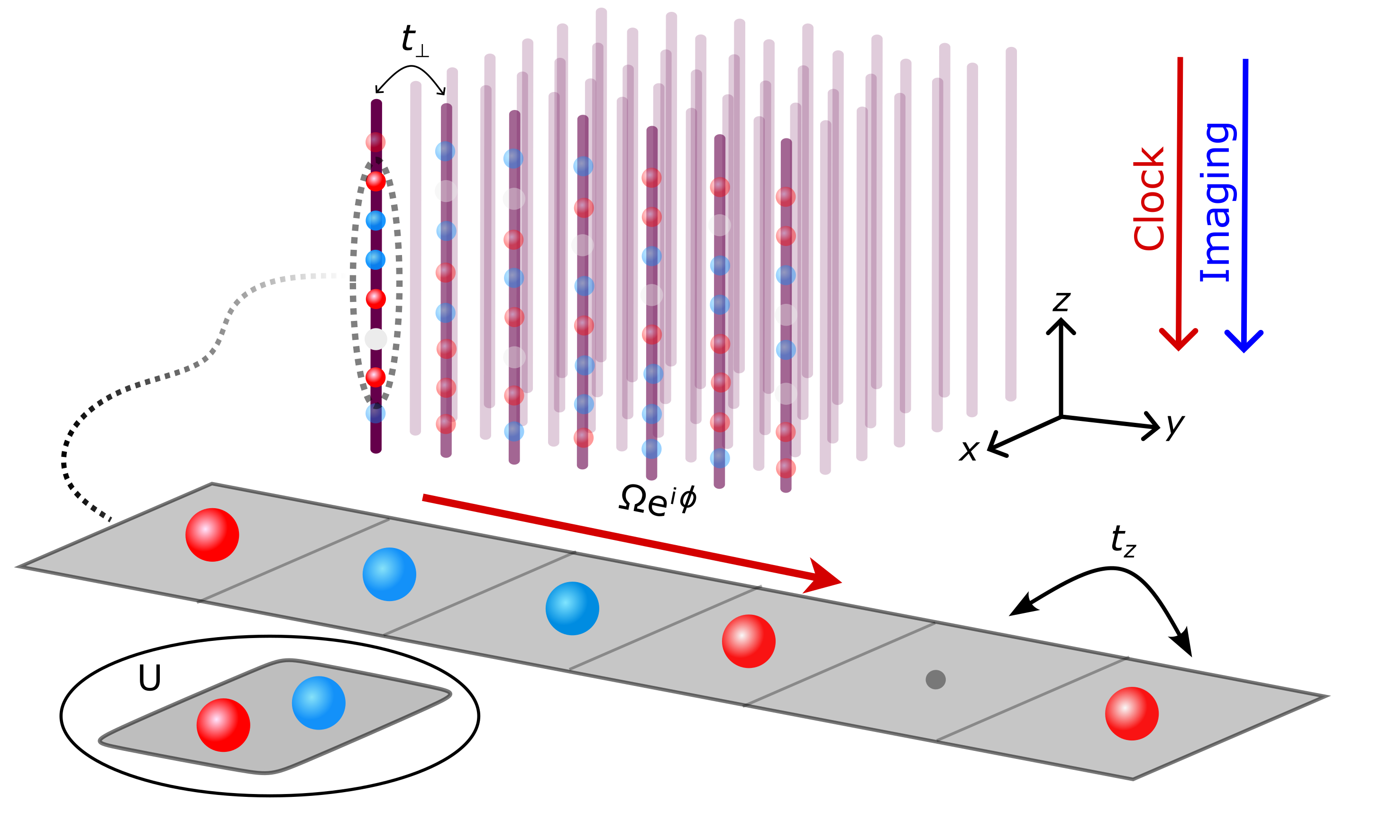}
\caption{\textbf{Three-dimensional optical lattice clock}. Atoms are isolated into 1D tubes by setting $t_{z} \gg t_{\perp}$. Introducing a clock drive along the $z$ spatial dimension, the modified Fermi-Hubbard Hamiltonian in Eq.~\ref{eq_FermiHubbard} is realized. Blue (red) spheres represent atoms in the ground (excited) states. The clock laser imprints a site-by-site spin-orbit-coupled phase $\phi$. 
}
\label{fig_schematic}
\end{figure*}

A schematic of the three-dimensional optical lattice clock studied in this work is presented in Fig.~\ref{fig_schematic}. The lattice has three Cartesian directions $x$,$y$,$z$ with independent lattice depths $V_x$, $V_y$, $V_z$. We assume the horizontal depths are equal $V_{x}=V_{y} \equiv V_{\perp}$, and prepare the system in an effectively one dimensional geometry by tuning $V_{\perp} \gg V_z$, so that tunneling takes place predominantly along the vertical direction $z$. The lattice is populated by fermionic atoms with two optically coupled internal spin states $e$, $g$, treated as a spin-1/2 degree of freedom. The clock laser is assumed to propagate along the $z$ direction. The atoms are also imaged along the $z$ direction for projective measurement, meaning that any experimental observable is integrated over all $z$. A more detailed technical overview of the experimental apparatus is provided in Ref.~\cite{milner_science}.

Since tunneling happens predominantly along $z$, the dynamics of atoms prepared in the lowest motional band of each independent tube along $z$ are captured by a 1D Fermi-Hubbard model,
\begin{equation}
\begin{aligned}
\hat{H}_{\mathrm{Hubbard}} = &- t_z \sum_{j}\sum_{\sigma = e,g} \left(\hat{c}_{j,\sigma}^{\dagger}\hat{c}_{j+1,\sigma} + h.c.\right) \\
&+ U \sum_{j}\hat{n}_{j,e}\hat{n}_{j,g}.
\end{aligned}
\end{equation}
Here $\hat{c}_{j,\sigma}$ annihilates an atom of spin $\sigma \in \{e,g\}$ on site $j$, $\hat{n}_{j,\sigma} = \hat{c}_{j,\sigma}^{\dagger}\hat{c}_{j,\sigma}$, $t_z$ is the tunneling rate along $z$, and $U$ is the on-site Hubbard repulsion. We always assume open boundary conditions unless otherwise specified. 

Due to the curvature of the beams forming the optical lattice, there is also an overall trapping potential. At the center of the trap, this potential is approximately harmonic:
\begin{equation}
\hat{H}_{\mathrm{trap}} = \eta_z \sum_{j} (j-j_0)^2 \left(\hat{n}_{j,e} + \hat{n}_{j,g}\right).
\end{equation}
The trap energy $\eta_z$ is set by the beam waists $W_x$, $W_y$ and depths $V_x$, $V_y$ of the lattice $x$, $y$ beams, and it is given by $\eta_z = \frac{2 V_x a^2}{W_x^2}+\frac{2 V_y a^2}{W_y^2}$. Here $j_0$ is the lattice site at the center of the harmonic potential. There is also a linear gravitational potential along $z$, which is incorporated by adjusting $j_0$; see Appendix~\ref{app_Potential} for further details on the confinement.

The final term in the Hamiltonian is the clock laser, given by,
\begin{equation}
\hat{H}_{\mathrm{laser}} = \frac{-i\Omega}{2} \sum_{j} \left(e^{i j \phi} \hat{c}_{j,e}^{\dagger}\hat{c}_{j,g} - h.c.\right).
\end{equation}
This term drives on-site spin-flips at a Rabi frequency $\Omega$. The laser introduces a travelling-wave phase factor $e^{i \vec{k}_{clk}\cdot \vec{r}}$ for laser wavevector $\vec{k}_{clk}$ and position $\vec{r}$. Since we consider a laser pointing along $z$, when the phase $\vec{k}_{clk}\cdot \vec{r}$ is projected onto the $z$ lattice direction, it yields a site dependent phase shift $\phi = |\vec{k}_c|a$ of the Rabi coupling rotation axis in the equatorial plane of the Bloch sphere, where $a$ is the lattice spacing. This site dependent phase shift generically breaks the SU(2) symmetry of the system and generates spin-spiral textures when the Rabi drive is applied. Specifically, we generate a spin-spiral with phase $\phi$ per lattice site.

It is convenient to make a basis transformation that transfers the spin-orbit phase away from the clock laser, by defining new fermionic operators,
\begin{equation}
\hat{c}_{j,e} \to e^{i j \phi/2}\hat{c}_{j,e}, \>\>\>\hat{c}_{j,g} \to e^{-i j \phi/2}\hat{c}_{j,g}.
\end{equation}
Under this transformation, the clock laser Hamiltonian becomes translationally invariant,
\begin{equation}
\hat{H}_{\mathrm{laser}} \to \hat{H}_{\mathrm{laser}}^{'} = \frac{-i\Omega}{2}\sum_j \left(\hat{c}_{j,e}^{\dagger}\hat{c}_{j,g} - h.c.\right).
\end{equation}
Hereafter, a prime superscript indicates Hamiltonian in this new rotated frame. The Fermi-Hubbard model picks up the spin-orbit phase,
\begin{equation}
\begin{aligned}
\label{eq_FermiHubbard}
\hat{H}_{\mathrm{Hubbard}}' = &- t_z \sum_{j}\left(e^{\frac{i \phi}{2}}\hat{c}_{j,e}^{\dagger}\hat{c}_{j+1,e} +e^{-\frac{i \phi}{2}}\hat{c}_{j,g}^{\dagger}\hat{c}_{j+1,g}+ h.c.\right)\\
&+ U \sum_{j}\hat{n}_{j,e}\hat{n}_{j,g},
\end{aligned}
\end{equation}
while the trapping potential remains invariant with $\hat{H}_{\mathrm{trap}}' = \hat{H}_{\mathrm{trap}}$.

Frequency shifts as well as the Ramsey fringe contrast can be experimentally probed by the imaging spectroscopy technique presented in Ref.~\cite{marti2018imaging}. In a spatial region of the cloud $r$, the excitation fraction $P(r) =  1/2 + \frac{C}{2}\cos(2 \pi f(r) t + \varphi_{0})$, where $f(r)$ is the local clock frequency, $C$ is the Ramsey fringe contrast, $t$ is the spectroscopy dark time, and $\varphi_{0}$ is a phase offset. $P(r)$ is experimentally determined via absorption imaging of the density distribution for atoms in the ground and metastable clock state $N_{e}(r)$ and $N_{g}(r)$ to determine the local excitation fraction $P(r) = N_{e}(r) / ( N_{e}(r) +  N_{g}(r))$. The clock laser phase is common-mode to all atoms in the ensemble, enabling probing \textit{differential} frequency shifts at dark times far surpassing the coherence time of the clock laser to directly determine the atomic coherence time of the atom ensemble. Although this enables resolving frequency shifts transverse to the imaging axis, frequency shifts along the imaging axis are integrated.

%%%%%
\section{Superexchange interactions}\label{superex_sec}
%%%%%

We now consider the regime of strong onsite repulsion $U \gg t_{z}$. In conventional studies of the Hubbard model, the low-energy physics may be described by an effective superexchange spin-1/2 model when at half filling (one atom per site). The spin $\ket{\uparrow'}_j$ and $\ket{\downarrow'}_j$ states are identified with a single atom of spin $e$ and $g$ on site $j$ respectively; these spin states are in the frame where the laser's spin-orbit phase has been gauged away. We assume that strong repulsion causes double occupancies to be energetically forbidden. To leading order, the tunneling mediates spin superexchange interactions between neighbouring atoms, which can be derived via standard second-order degenerate perturbation theory. For a chain of $L$ filled sites with open boundary conditions, these interactions read (see Appendix~\ref{app_Superexchange} for derivation):
\begin{equation}
\label{eq_Superexchange}
\hat{H}' = \sum_{j=1}^{L-1} V_j \left[\frac{1}{2}\left(e^{i \phi} \hat{s}_j^{+} \hat{s}_{j+1}^{-} + h.c.\right) + \hat{s}_j^z \hat{s}_{j+1}^z\right] + \Omega \sum_j \hat{s}_j^y,
\end{equation}
where $\hat{s}_j^{x}=\frac{1}{2}\left(\ket{\uparrow'}_j\bra{\downarrow'}_j + h.c.\right)$, $\hat{s}_j^{y}=\frac{-i}{2}\left(\ket{\uparrow'}_j\bra{\downarrow'}_j - h.c.\right)$, $\hat{s}_j^{z}=\frac{1}{2}\left(\ket{\uparrow'}_j\bra{\uparrow'}_j -\ket{\downarrow'}_j\bra{\downarrow'}_j \right)$. The interaction coefficient is:
\begin{equation}
\label{eq_superexchangeStrength}
V_j = \frac{4t_z^2 U}{U^2 - [2(j-j_0)+1]^2 \eta_z^2}.
\end{equation}
The factor $[2(j-j_0)+1]\eta_z$ is the local potential difference between sites $j$ and $j+1$, scaling linearly with site index. The model is valid in the regime where $|U \pm [2(j-j_0)+1]\eta_z| \gg t_z$ for all sites $j$. We also assume that the Rabi frequency is not too strong compared to the on-site interaction $|\Omega| \ll U$ (provided the laser is turned on with $\Omega \neq 0$) as otherwise it will shift the energies in the superexchange denominators. A more general treatment is provided in Ref.~\cite{mamaev2021tunable}. 

We note that there can be parameters for which the superexchange denominators become too small, $|U \pm [2(j-j_0)+1]\eta_z| \lesssim t_z$, for a few lattice sites, which causes perturbation theory to break down, yielding resonant physics beyond the spin model~\cite{mamaev2022resonant, bukov2016schrieffer, mamaev2019quantum, xu2018correlated}. We nonetheless anticipate that in the presence of strong harmonic confinement, such physics will remain localized to the few sites where perturbation theory fails, and the effective spin model will remain a valid description for the vast majority of atoms in the cloud.

We now discuss the effects of the initial lattice filling fraction. When the temperature is optimized for a spin polarized Fermi gas, recent ultracold atom experiments have been able to achieve filling on the order of $\sim 90\%$ in the center of the 3D optical lattice~\cite{milner2023high, mukherjee2017homogeneous, omran2015microscopic}. Achieving close to unity filling over large spatial regions of the cloud is challenging, due to both stringent temperature requirements and trapping potential inhomogeneities. In general, for Hubbard-like models even $\sim 10\%$ holes can have a strong effect on the dynamics since the bare tunneling rate $t_z$ is much stronger than the perturbative superexchange rate $V_j$. However, it is crucial to note that even for fillings below one atom per site, if the trapping potential energy difference $[2(j-j_0)+1]\eta_z$ between two given sites is large compared to $t_z$, then direct atomic motion into empty sites is suppressed while spin dynamics can still occur at rates $\approx V_j$~\cite{dimitrova2020enhanced}. Since the potential difference scales linearly with site index $j$, if the trap energy $\eta_z$ is at least comparable to the tunneling rate $t_z$, then for all but a few sites at the center of the trap, atoms will not be able to tunnel into adjacent holes. We thus approximate that a vertical tube along $z$ can be split into stationary independent uninterrupted chains of atoms, as depicted in Fig.~\ref{fig_ContrastChains}(a). Empty sites act as boundaries between these chains. The dynamics of the whole tube is determined by solving each chain independently, and summing their contributions to a desired observable such as spin contrast. This approximation will be benchmarked further in Section~\ref{sec_Holes}.

A key parameter controlling the superexchange dynamics is thus the length, $L$, of an \textit{uninterrupted} chain (all sites populated  along  the chain) within a tube following the lattice loading process.  We assume that the loading results in an initial distribution of atoms and holes modeled by a Fermi-Dirac distribution with an effective temperature $T$, relative to the Fermi temperature $T_F$. Appendix~\ref{app_Filling} describes this model and the subsequent calculations in detail. Fig.~\ref{fig_ContrastChains}(b) shows the full distribution of initial atomic density $\langle\hat{n}_j\rangle$ (probability of finding a single atom at site $j$) within the tube for different temperatures. Fig~\ref{fig_ContrastChains}(c) shows how many uninterrupted chains $N_L$ of length $L$ one may find on average. A hot gas with $T/T_F \gg 1$ mostly has isolated atoms $L=1$ with a few pairs $L=2$, fewer three-site chains $L=3$ and so on, thus $N_{1} \gg N_{2} \gg N_{3}$ etc. In the opposite limit of a cold gas $T \to 0$, there will only be a single chain at the center of the trap with length equal to the number of atoms in the tube. This analysis assumes that the chemical potential is still below the band-gap of the lattice for all directions, as otherwise motionally excited states at the center of the trap would become populated before ground-band states far from the center.

%%%%%
\section{Ramsey contrast decay}
\label{ramsey_sec}
%%%%%

We now provide a theoretical description of typical dynamics expected in a 3D optical lattice. Specifically, we study contrast decay in Ramsey spectroscopy. 
We first assume that each uninterrupted chain of atoms starts spin-polarized in the ground state $\prod_{j=1}^{L} \ket{\downarrow'}_j$. The atoms are then laser driven with a strong Rabi frequency $\Omega \gg U, t_z$, such that $\hat{H}_{\mathrm{laser}}$ is the only relevant term in the Hamiltonian. Driving for a time $t_{\mathrm{init}}\Omega = \pi/2$ implements a $\pi/2$ pulse and generates a product state with each atom in a superposition state,
\begin{equation}
\ket{\psi(0)}_x=\prod_{j=1}^{L} \frac{1}{\sqrt{2}}\left(\ket{\uparrow'}_j + \ket{\downarrow'}_j\right).
\end{equation}
After preparing this state, the laser is turned off ($\Omega = 0$) and the state evolves freely under the spin Hamiltonian $\hat{H}'|_{\Omega = 0}$ for a dark time $t$. The spin contrast for a chain of $L$ sites is then obtained by measuring,
\begin{equation}
\label{eq_Contrast}
C_L(t) = 
\frac{2}{L}\sqrt{\langle\hat{S}_L^{x}(t)\rangle^2 + \langle\hat{S}_L^{y}(t)\rangle^2},
\end{equation}
where $\langle\hat{S}_L^{\alpha}(t)\rangle$ is the time-evolved collective spin observable $\hat{S}^{\alpha} = \sum_{j=1}^{L}\hat{s}_j^{\alpha}$ given by $\langle\hat{S}_L^{\alpha}(t)\rangle = \bra{\psi(0)}_x e^{i t \hat{H}'|_{\Omega=0}}\hat{S}^{\alpha} e^{-i t \hat{H}'|_{\Omega=0}}\ket{\psi(0)}_x$. 

\begin{figure*}
\center
\includegraphics[width=1\textwidth]{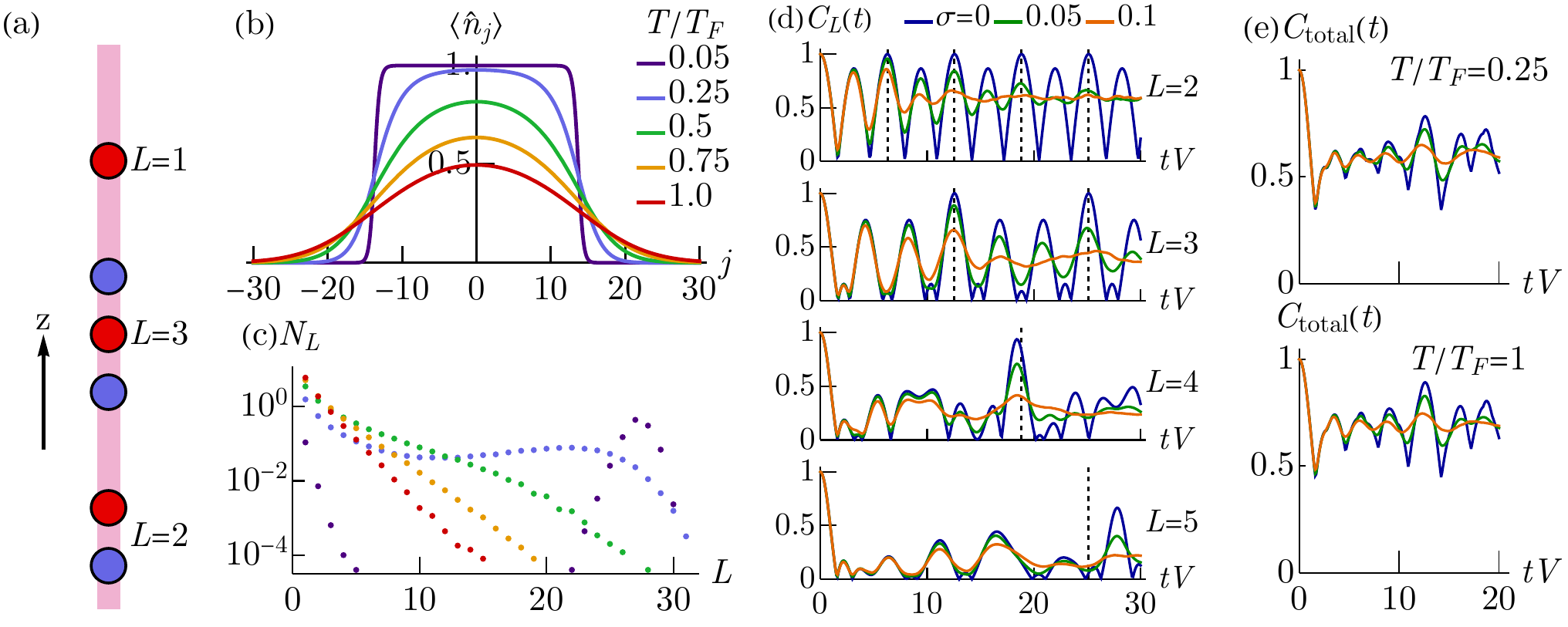}
\caption{\textbf{Ramsey spectroscopy of small spin chains in the presence of inhomogeneity}. (a) Depiction of a 1D tube along the vertical $z$ lattice direction filled by atoms following state preparation. Empty sites act as walls, as the harmonic trap inhibits direct tunneling. Uninterrupted chains of $L$ atoms undergo spin-1/2 superexchange interactions. (b) Fermi-Dirac distribution of atom density in the tube. (c) Average number $N_L$ of isolated chains of length $L$ (padded by empty sites) in a single 1D tube at the horizontal center of the trap, for a specific total atom number (see Appendix~\ref{app_Filling}). (d) Superexchange-induced contrast dynamics [Eq.~\eqref{eq_Contrast}] for isolated chains of length $L$, for different levels of disorder $\sigma$ (in \%) in the nearest-neighbour superexchange coupling $V_j$. Rather than averaging over an explicit set of couplings $V_j$ based on the experimental harmonic confinement, we instead draw from a generic Gaussian distribution of mean $V$ and width $\sigma V$. Vertical dashed lines are integer multiples of $tV=2\pi (L-1)$. (e) Dynamics of total contrast for a 1D tube at the horizontal center of the trap, obtained by averaging the contributions of isolated chains of different lengths according to the distribution in panel (c) [via Eq.~\eqref{eq_TotalContrast}]. We only add contributions from chains of length $L = 1 \dots 16$, assuming the rest to be negligible for these temperatures. We still use a generic Gaussian randomly sampled disorder in $V_j$.
}
\label{fig_ContrastChains}
\end{figure*}

It is important to observe that even before we account for inhomogeneities in the superexchange couplings $V_j$, different chain lengths $L$ already undergo dramatically different contrast dynamics under the above protocol. Length $L=1$ (i.e. isolated atoms) has no contrast decay under this model with $C_{L=1}(t)= 1$. For $L=2$ sites, the contrast takes a simple form,
\begin{equation}
C_{L=2}(t)= \left|\cos^2\left(\frac{\phi}{2}\right)+ \sin^2\left(\frac{\phi}{2}\right)\cos\left(V t\right)\right|,
\end{equation}
with $V_j=V$ (there is only one coupling for two sites). The contrast exhibits oscillations with a single frequency set by the coupling strength $V$ and amplitude determined by the SOC phase. For $L=3$ sites and beyond there are multiple oscillating frequencies, but an analytic description quickly grows intractable; Appendix~\ref{app_Analytics} provides a few results.

While general analytics are intractable, short length chains are readily solved by exact numerical time-evolution. With numerics, we can readily incorporate the effects of inhomogeneous interaction coefficients $V_j$ as well. These coefficients vary significantly due to the lattice site dependence on $j$ in Eq.~\eqref{eq_superexchangeStrength}, the overall vertical position of each chain (i.e. the corresponding trap center $j_0$ for each chain separately). In a real experimental setup, there can also be inhomogeneities in the Hubbard $t_z$ and $U$ parameters across the lattice. In principle one can perform a careful averaging over all experimental conditions. However, such averaging can be very numerically expensive. If we are just interested in the collective spin contrast, much of these inhomogeneities can be made more abstract by drawing $V_j$ from a generic probability distribution, and averaging over that. In what follows, we take $V_j$ from a Gaussian distribution of mean $V$ and standard deviation $\sigma V$; i.e. $\sigma = 0.1$ corresponds to $10 \%$ standard deviation in the coupling strengths.

Fig~\ref{fig_ContrastChains}(d) plots numerically-computed contrast decay for different small chain lengths $L$. Disorder $\sigma$ in the couplings generally leads to a damping of oscillations at longer dark times. However, short chains exhibit a revival of contrast at dark time $tV= 2 \pi (L-1)$, scaling with the length of the chain $L$. Analogous revivals can be found in Ising spin models~\cite{foss2013nonequilibrium}.

Since these revivals do not occur at the same time for different chain lengths, the Ramsey signal for the whole lattice will exhibit a sharper peak at times when multiple chain lengths' revivals coincide.  We can see this effect by considering the total contrast of an entire vertical 1D tube for different initial temperatures (hence filling fractions, and lengths of uninterrupted chains). Specifically, we write the total contrast as,
\begin{equation}
\label{eq_TotalContrast}
C_{\mathrm{total}}(t) = \frac{2}{\sum_{L=1}^{\infty}N_L L}\sqrt{\langle\hat{S}^x_{\mathrm{total}}(t)\rangle^2 + \langle\hat{S}^y_{\mathrm{total}}(t)\rangle^2},
\end{equation}
where $N_L$ is the average number of chains of length $L$ in a tube from Fig.~\ref{fig_ContrastChains}(c), and $\langle\hat{S}^{\alpha}_{\mathrm{total}}(t)\rangle$ are collective spin observables given by,
\begin{equation}
\langle\hat{S}^{\alpha}_{\mathrm{total}}(t)\rangle = \sum_{L=1}^{\infty} N_L \langle\hat{S}_{L}^{\alpha}(t)\rangle,
\end{equation}
with $\langle\hat{S}_{L}^{\alpha}(t)\rangle$ the time-evolved collective spin observables for a chain of length $L$ as before.

In Fig.~\ref{fig_ContrastChains}(e) we plot the total contrast for different initial temperatures, still modelling disorder in the couplings $V_j$ by drawing from a generic Gaussian distribution. A stronger revival of contrast occurs at dark time $t V = 4\pi$, where chains of length $L =2$ and $L=3$ both exhibit a revival. We also find that the overall contrast decay profile remains relatively robust independent of temperature. A hotter gas generally has more isolated $L=1$ atoms, which do not contribute any oscillatory dynamics, but simply raise/lower the long-time average of the signal since they have constant $C_{L=1}(t) = 1$ under the model. Short chains of length $L=2,3$ undergo oscillatory dynamics, which average out to a non-zero mean contrast when the interaction strengths (hence the oscillation period) are disordered. One only starts to see qualitative differences at very low temperatures, where the dynamics are dominated by a single long chain. We have not plotted the result for very low $T/T_F$ since the resulting chain lengths are too long to numerically solve, but we predict the resulting contrast will simply decay to zero with no significant oscillatory dynamics. Contrast decay dynamics thus provides both a way to benchmark the strength of superexchange interactions (via e.g. Fourier analysis of the decay profile), and a means of thermometry (by studying the long-time average of $C_{\mathrm{total}}(t)$ to calibrate the initial temperature of the gas).

%%%%%
\section{Rabi spectroscopy}
\label{rabi_sec}
%%%%%

In optical lattice clocks, another important dynamical sequence is Rabi spectroscopy, which subjects the atoms to a continuous Rabi drive rather than just a single initial pulse. In this case, we consider time-evolution with each chain starting from an initial state:
\begin{equation}
\ket{\psi(0)}_z=\prod_{j=1}^{L}\ket{\downarrow'}_j.
\end{equation}
This state is evolved for a time $t$ under the full spin interaction $e^{-it \hat{H}'}\ket{\psi(0)}_z$, now with a finite Rabi frequency $\Omega \neq 0$. The final measurement is of the excitation fraction,
\begin{equation}
\hat{n}_{L}^{e} = \sum_{j=1}^{L} \left(\hat{s}_j^z + \frac{1}{2}\right),
\end{equation}
which we time-evolve as before via $\langle\hat{n}_L^e (t)\rangle = \bra{\psi(0)}_z e^{i t \hat{H}'} \hat{n}_L^{e} e^{-i t \hat{H}'} \ket{\psi(0)}_z$.

\begin{figure}
\center
\includegraphics[width=1.0\columnwidth]{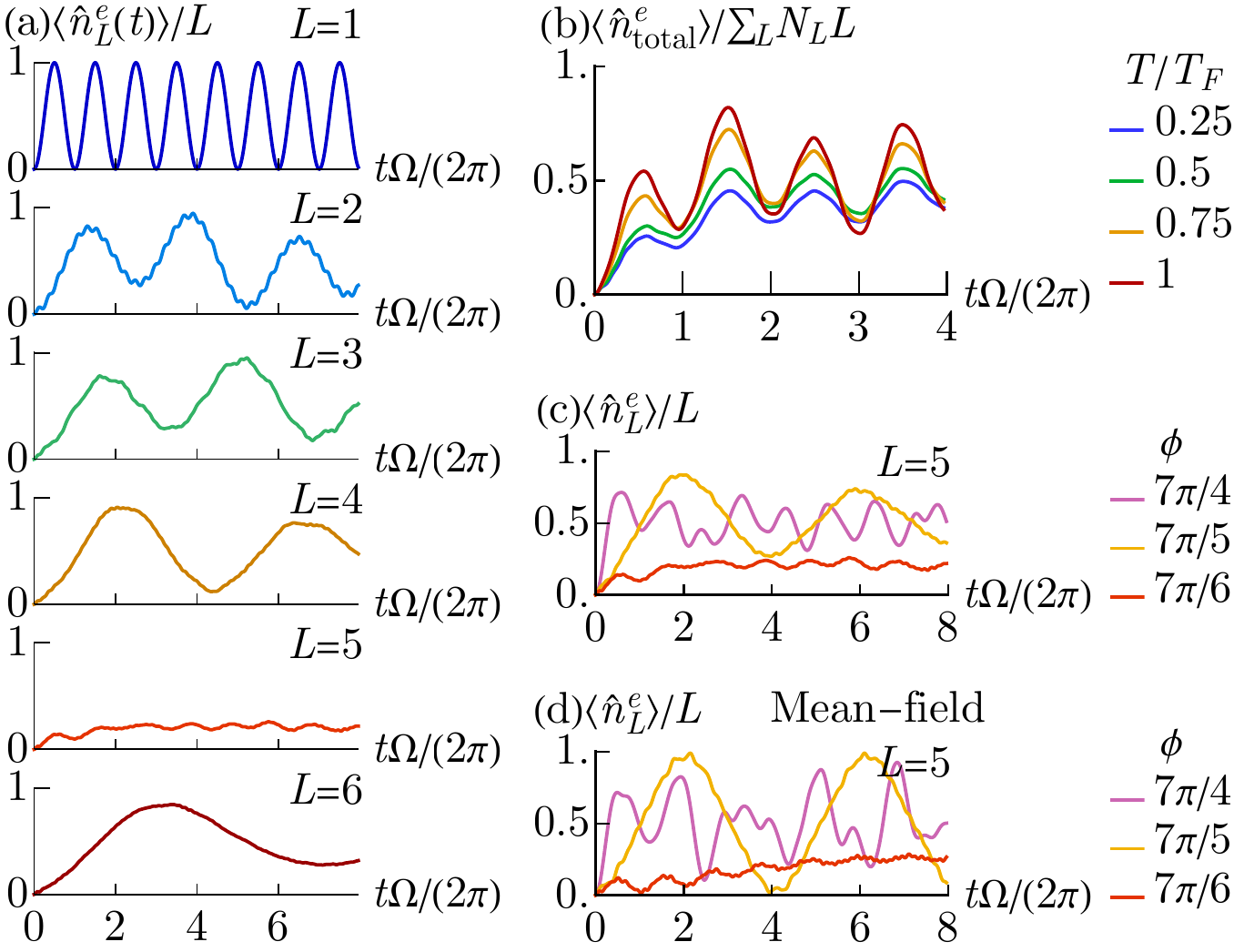}
\caption{\textbf{Rabi spectroscopy of small spin chains}. (a) Rabi evolution dynamics for superexchange-interacting chains of length $L$, using the Hamiltonian in Eq.~\eqref{eq_Superexchange} with homogeneous couplings $V_j = V$. The Rabi frequency is set to $\Omega/V = 0.25$, and the spin-orbit phase to $\phi = 7\pi/6$. (b) Rabi evolution averaged over different chain lengths, according to the thermal distribution in Fig.~\ref{fig_ContrastChains}(c). (c) Rabi evolution dynamics for fixed chain length $L=5$ and different spin-orbit phases. (d) Rabi evolution dynamics for the same parameters as panel (c), but solved numerically at the mean-field level (see Appendix~\ref{app_MeanField} for the equations of motion).
}
\label{fig_Rabi}
\end{figure}

For a strong Rabi frequency $\Omega \gg U, t_z$, the excitation fraction would simply undergo coherent oscillations $\langle \hat{n}_{L}^{e}\rangle = L \sin^2\left(\frac{\Omega t}{2}\right)$. A weak Rabi frequency $\Omega \ll V_j$ leads to qualitatively different physics, as the strongest energy scale is now set by the interactions themselves. Fig~\ref{fig_Rabi}(a) shows numerically-computed time-evolution of excitation fraction for weakly Rabi-driven isolated chains of different lengths $L$, assuming homogeneous superexchange couplings $V_j = V$ for simplicity (we anticipate disorder $\sigma$ to wash out long-time oscillatory features as before). We find that the maxima in the excitation fraction depend very sensitively on the length $L$ and the spin-orbit phase $\phi$.

As we did for Ramsey contrast decay, we can also define a total excitation fraction averaged over different chain lengths,
\begin{equation}
\langle\hat{n}_{\mathrm{total}}^{e}(t)\rangle = \sum_{L=1}^{\infty} N_L \langle\hat{n}_{L}^e (t)\rangle,
\end{equation}
In Fig.~\ref{fig_Rabi}(b) we show this total excitation fraction. Even for fairly cold initial temperatures, one can find a higher excitation fraction at the second Rabi peak $t \Omega = 3 \pi$ (rather than the first $t \Omega = \pi$). Physically, this occurs because the peak excitation fraction of short $L = 2, 3, 4$ chains are shifted to later times due to the interactions, and such short chains comprise a significant fraction of the atomic population even for cold temperatures $T / T_F \ll 1$. In the true limit of zero temperature $T / T_F \to 0$ where the initial condition is a single long uninterrupted chain, we do not anticipate any oscillations, save for possible revivals on timescales $\sim 1/L$ for homogeneous couplings.

We also comment that the position of the first peak as a function of chain length $L$ and spin-orbit phase $\phi$ exhibits a complex, highly non-monotonic behavior. As seen in Fig.~\ref{fig_Rabi}(a), going from chain length $L =4$ to 5 to 6 yields dramatic changes from coherent oscillation to an overall lack of signal. This non-monotonicity can be attributed to the interplay of the underlying spin-spiral texture imprinted by the clock laser, the anisotropic spin interactions, and the boundary conditions of the open length-$L$ chain. Fig.~\ref{fig_Rabi}(c) shows the dynamics for a fixed chain length, but varied SOC phase $\phi$, showcasing the dramatic change. While predicting the outcome for a generic $L$ and $\phi$ is challenging, the qualitative behavior can be captured on a mean-field level (see Appendix~\ref{app_MeanField} for the equations). Fig.~\ref{fig_Rabi}(d) shows the same time-evolution at the mean-field level, which does not quantitatively match the full exact numerical simulations done thus far, but reproduces the general trend.

While non-trivial dynamics as a function of chain length $L$ can pose a challenge for Rabi spectroscopy, it may also offer a tool for controlling the quantum dynamics of the system. For instance, suppose one has a hot gas with mostly chains of length $L=1,2,3$. One may evolve to a Rabi time $t\Omega$ for which chains of length $L=2$ exhibit near-maximum excitation fraction while $L=3$ are at a minimum, then use resonant fluorescence to remove all excited atoms - thereby getting rid of the $L=2$ chains while keeping the $L=3$ ones. Doing this allows one to study non-monotonic physics from $L=3$ chains only (or vice versa). With some more careful optimization of parameters, or multiple dynamical sequences, longer chains could also be isolated. This offers a tool for studying fixed-length chain dynamics without requiring site-resolved spatial resolution.

%%%%%
\section{Imperfect filling fraction}
\label{sec_Holes}
%%%%%

Thus far, we have assumed that if there are holes in the initial filling fraction, they are localized by the harmonic trapping potential, and act as walls for independent spin chains. This approximation breaks down when the tunneling rate $t_z$ becomes comparable to the local energy differences $[2(j-j_0)+1]\eta_z$. If we still have a strong on-site Hubbard repulsion $U \gg t_z$, doubly-occupied sites are still energetically forbidden. In this case, we can describe the low-energy physics of an entire vertical 1D tube with a spin-1 $t-J$ model rather than just isolated spin-1/2 chains. Each lattice site is described by a spin-1 degree of freedom; states $\ket{+1}_j$, $\ket{-1}_j$ correspond to an $e$, $g$ atom on that site respectively, while $\ket{0}_j$ means the site is empty. Degenerate second-order perturbation theory yields the following effective spin-1 model (see Appendix~\ref{app_Superexchange} for details):
\begin{widetext}
\begin{equation}
\begin{aligned}
\label{eq_tJ}
\hat{H}'_{\mathrm{Spin-1}} &= -t_z \sum_{j}\left(e^{i \phi/2} \hat{S}_{j}^{+}\hat{S}_{j+1}^{-} +h.c.\right)\hat{S}_{j}^{z}\hat{S}_{j+1}^{z}+\frac{t_z}{2}\sum_{j}\left(e^{i \phi/2} \hat{S}_{j}^{+}\hat{S}_{j+1}^{-} - h.c.\right)\left(\hat{S}_{j}^{z} - \hat{S}_{j+1}^{z}\right)\\
&+\sum_{j}4V_j\big(\frac{e^{i \phi}}{2}\hat{S}_{j}^{+}\hat{S}_{j}^{+}\hat{S}_{j+1}^{-}\hat{S}_{j+1}^{-}+\frac{e^{-i \phi}}{2}\hat{S}_{j}^{-}\hat{S}_{j}^{-}\hat{S}_{j+1}^{+}\hat{S}_{j+1}^{+}+\hat{S}_{j}^{z}\hat{S}_{j+1}^{z} - \hat{S}_{j}^{z}\hat{S}_{j}^{z} \hat{S}_{j+1}^{z}\hat{S}_{j+1}^z\big).
\end{aligned}
\end{equation}
\end{widetext}
Here $\hat{S}_j^{+}$, $\hat{S}_j^{-}$, $\hat{S}_{j}^{z}$ are spin-1 operators. The first line captures direct hopping of atoms; the flip-flop interaction $\hat{S}_j^{+}\hat{S}_{j+1}^{-}$ swaps a particle (spin states $\ket{+1}$ or $\ket{-1}$) with a hole (spin state $\ket{0}$), and does nothing otherwise. The terms proportional to $\hat{S}_j^{z}$, $\hat{S}_{j+1}^z$ ensure the correct fermionic signs of the hopping processes are respected. Note that this model is only robust in 1D. For 2D and above, the hopping matrix elements would pick up fermionic strings, which can still be modeled with a spin-1 description but become complex non-local operators. The second line is the superexchange interactions from before, with the same coefficient $V_j$. Terms like $\hat{S}_j^{+}\hat{S}_j^{+}\hat{S}_{j+1}^{-}\hat{S}_{j+1}^{-}$ encode spin flip-flop interactions exchanging $\ket{+1}$ and $\ket{-1}$, while the remaining diagonal terms involving $\hat{S}_{j}^{z}$, $\hat{S}_{j
+1}^{z}$ encode the Ising portion of the Hamiltonian. Our model omits interaction-mediated double hops, such as an atom hopping over its neighbour and into a further empty site, e.g. from $\ket{g,e,0}$ to $\ket{0,e,g}$ in a three-site chain. While such terms may also be computed perturbatively, we have found that they do not contribute a significant correction to our results. We also omitted the Rabi drive - if it is present, we would have an additional Hamiltonian term $\frac{-i\Omega}{4}\sum_j \left(\hat{S}_j^+ \hat{S}_j^+ -h.c.\right)$.

\begin{figure}
\center
\includegraphics[width=0.85\columnwidth]{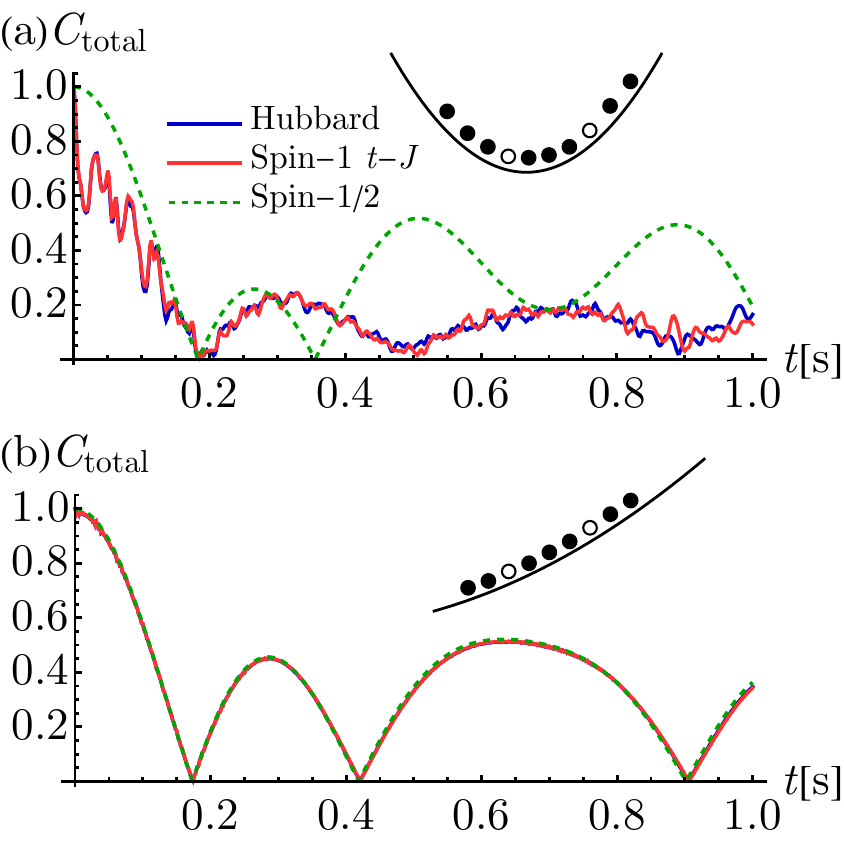}
\caption{\textbf{Benchmarking spin models under realistic parameters.} (a) Ramsey spectroscopy contrast dynamics for a distribution of particles (shown in inset) close to the center of the harmonic trap, modeled by a Fermi-Hubbard description [Eq.~\eqref{eq_FermiHubbard}], a spin-1 $t-J$ model [Eq.~\eqref{eq_tJ}], and a spin-1/2 model for isolated chains [Eq.~\eqref{eq_Superexchange}]. The latter approach solves the dynamics of each uninterrupted chain separately and sums the contributions together. We use parameters of $t_z/h=23$ Hz, $\eta_z/h = 17$ Hz, $U/h=1.4$ kHz (corresponding to lattice depths $V_x=V_y=30 E_R$, $V_z = 15 E_r$). The initial condition is depicted in the inset; solid dots are atoms starting in $\ket{\uparrow'}_j + (\ket{\downarrow'}_j)/\sqrt{2}$, empty circles are vacancies, and the parabola is the harmonic potential profile. (b) Contrast dynamics for a distribution of particles (inset) far from the trap center.
}
\label{fig_Holes}
\end{figure}

We validate this model in Fig.~\ref{fig_Holes}, which plots contrast decay under Ramsey spectroscopy. As anticipated, we find that for a set of atoms close to the center of the trap, the naive prediction of isolated spin-1/2 chains breaks down, while the spin-1 description remains valid. The rapid decay in contrast at short time occurs due to direct atomic motion at the bottom of the trap, where the potential is not enough to suppress it. Contrast does not fall to zero as the atoms further from the trap center are still somewhat localized. For such atoms, both the spin-1/2 and spin-1 descriptions are in good agreement, rendering the latter unnecessary. Provided the filled lattice is wide enough, most atoms will fall into the latter regime, and thus our prior predictions with spin-1/2 chains will remain valid without having to solve the more computationally complex spin-1 model. The spin-1 mapping is nonetheless useful, as it already provides a better Hilbert space size scaling for exact simulations. One could also more easily apply approximate numerical techniques such as the discrete truncated Wigner approximation (DTWA) to spin interactions~\cite{schachenmayer2015many}, which may be useful to capture the physics in situations where the transverse tunneling $t_{\perp}$ becomes non-negligible (meaning the system is 3D, where exact numerical treatments are intractable).

%%%%%
\section{Conclusions and outlook}\label{conclusions_sec}
%%%%%

We have shown that the Ramsey and Rabi signals measured by 3D optical lattice clocks can see significant effects from superexchange interactions. For currently accessible gas temperatures, these effects stem from competing contributions of short interacting spin chains of different lengths, which feature vastly different length-dependent physics due to SOC imposed by the clock laser. In principle such effects can be mitigated by preparing a colder gas, operating in a deep lattice along the direction of the clock laser, or operating in a regime where $\phi \text{ mod } 2 \pi = 0$ via using an accordion lattice to modify the lattice spacing while still fulfilling the magic wavelength condition~\cite{hutson2019engineering}. Since the physics of individual short chains are easily solved (at least numerically), one can also employ techniques such as Rabi pulses on longer timescales $t\Omega = \pi, 2\pi$, etc., for which multiple short-length chains (e.g. $L=2$ and $L=3$) happen to exhibit revivals, yielding an improved spin contrast. In addition to contrast decay, superexchange interactions can also give rise to unwanted frequency shifts that can limit clock accuracy. However, for standard Ramsey spectroscopy protocols where $\langle \hat{S}^{z} \rangle = 0$, the associated frequency shift is zero at the mean-field level. Raman scattering can modify the spin projection, and was modeled in a one dimensional lattice clock~\cite{kim2025atomic}.

In-situ anisotropic spin interactions can also be a tool rather than a detriment. Such interactions can be used to generate metrologically useful quantum entanglement in the form of spin-squeezing~\cite{kitagawa1993squeezed, ma2011quantum}, which can further bolster the precision of the clock~\cite{he2019engineering, hernandez2022one}. Since different spin chain lengths will exhibit maximal squeezing at different evolution times, one can improve overall squeezing by optimizing over different length contributions, or employing dynamical decoupling protocols to ensure the maximal squeezing times coincide. There is some promise of leveraging the 3D lattice dimensionality to create a more strongly squeezed state~\cite{mamaev2024spin}.

\textit{Acknowledgments - } We thank Jun Ye, Stefan Lannig, Wolfgang Ketterle, and Matjaz Kebric for helpful discussions. W.R.M acknowledges funding from the MIT physics department in Wolfgang Ketterle's group.  AMR acknowledges support from AFOSR FA9550-24-1-0179, the NSF JILA-PFC PHY-2317149, and by NIST.

\bibliography{superexchangeBib}% Produces the bibliography via BibTeX.

\providecommand{\noopsort}[1]{}\providecommand{\singleletter}[1]{#1}%
\begin{thebibliography}{45}%
\makeatletter
\providecommand \@ifxundefined [1]{%
 \@ifx{#1\undefined}
}%
\providecommand \@ifnum [1]{%
 \ifnum #1\expandafter \@firstoftwo
 \else \expandafter \@secondoftwo
 \fi
}%
\providecommand \@ifx [1]{%
 \ifx #1\expandafter \@firstoftwo
 \else \expandafter \@secondoftwo
 \fi
}%
\providecommand \natexlab [1]{#1}%
\providecommand \enquote  [1]{``#1''}%
\providecommand \bibnamefont  [1]{#1}%
\providecommand \bibfnamefont [1]{#1}%
\providecommand \citenamefont [1]{#1}%
\providecommand \href@noop [0]{\@secondoftwo}%
\providecommand \href [0]{\begingroup \@sanitize@url \@href}%
\providecommand \@href[1]{\@@startlink{#1}\@@href}%
\providecommand \@@href[1]{\endgroup#1\@@endlink}%
\providecommand \@sanitize@url [0]{\catcode `\\12\catcode `\$12\catcode
  `\&12\catcode `\#12\catcode `\^12\catcode `\_12\catcode `\%12\relax}%
\providecommand \@@startlink[1]{}%
\providecommand \@@endlink[0]{}%
\providecommand \url  [0]{\begingroup\@sanitize@url \@url }%
\providecommand \@url [1]{\endgroup\@href {#1}{\urlprefix }}%
\providecommand \urlprefix  [0]{URL }%
\providecommand \Eprint [0]{\href }%
\providecommand \doibase [0]{https://doi.org/}%
\providecommand \selectlanguage [0]{\@gobble}%
\providecommand \bibinfo  [0]{\@secondoftwo}%
\providecommand \bibfield  [0]{\@secondoftwo}%
\providecommand \translation [1]{[#1]}%
\providecommand \BibitemOpen [0]{}%
\providecommand \bibitemStop [0]{}%
\providecommand \bibitemNoStop [0]{.\EOS\space}%
\providecommand \EOS [0]{\spacefactor3000\relax}%
\providecommand \BibitemShut  [1]{\csname bibitem#1\endcsname}%
\let\auto@bib@innerbib\@empty
%</preamble>
\bibitem [{mil(2025)}]{milner_science}%
  \BibitemOpen
  \href@noop {} {\bibinfo {title} {{W.~R. Milner, S. Lannig, M. Aeppli,L. Yan,
  A. Chu, B. Lewis, M.~N. Frankel, R.~B. Hutson, A.~M. Rey, and J. Ye,
  \textit{Coherent evolution of superexchange interaction in seconds long
  optical clock spectroscopy}}}} (\bibinfo {year}
  {\href{https://www.science.org/doi/10.1126/science.ado5987}{\textcolor{blue}{Science
  \textbf{388}, 503 (2025)}}})\BibitemShut {NoStop}%
\bibitem [{kol(2016)}]{kolkowitz2016gravitational}%
  \BibitemOpen
  \href@noop {} {\bibinfo {title} {{S. Kolkowitz, I. Pikovski, N. Langellier,
  M.~D. Lukin, R.~L. Walsworth, J. Ye, \textit{Gravitational wave detection
  with optical lattice atomic clocks}}}} (\bibinfo {year}
  {\href{https://journals.aps.org/prd/abstract/10.1103/PhysRevD.94.124043}{\textcolor{blue}{Phys.
  Rev. D \textbf{94}, 124043 (2016)}}})\BibitemShut {NoStop}%
\bibitem [{san(2019)}]{sanner_lorentz}%
  \BibitemOpen
  \href@noop {} {\bibinfo {title} {{C. Sanner, N. Huntemann, R. Lange, C. Tamm,
  E. Peik, M.~S. Safronova, and S.~G. Porsev, \textit{Optical clock comparison
  for Lorentz symmetry testing}}}} (\bibinfo {year}
  {\href{https://www.nature.com/articles/s41586-019-0972-2}{\textcolor{blue}{Nature
  (London) \textbf{567}, 204 (2019)}}})\BibitemShut {NoStop}%
\bibitem [{ken(2020)}]{kennedy2020precision}%
  \BibitemOpen
  \href@noop {} {\bibinfo {title} {{C.~J. Kennedy, E. Oelker, J.~M. Robinson,
  T. Bothwell, D. Kedar, W.~R. Milner, G.~E. Marti, A. Derevianko, and J. Ye,
  \textit{Precision Metrology Meets Cosmology: Improved Constraints on
  Ultralight Dark Matter from Atom-Cavity Frequency Comparisons}}}} (\bibinfo
  {year}
  {\href{https://journals.aps.org/prl/abstract/10.1103/PhysRevLett.125.201302}{\textcolor{blue}{Phys.
  Rev. Lett. \textbf{125}, 201302 (2020)}}})\BibitemShut {NoStop}%
\bibitem [{kat(2020)}]{katori_GR}%
  \BibitemOpen
  \href@noop {} {\bibinfo {title} {{M. Takamoto, I. Ushijima, N. Ohmae, T.
  Yahagi, K. Kokado, H. Shinkai, and H. Katori, \textit{Test of general
  relativity by a pair of transportable optical lattice clocks}}}} (\bibinfo
  {year}
  {\href{https://materias.df.uba.ar/l5a2022v/files/2022/02/TEP-Grupo-8-Relojes-opticos-transportables.pdf}{\textcolor{blue}{Nat.
  Photonics \textbf{14}, 411 (2020)}}})\BibitemShut {NoStop}%
\bibitem [{rmp(2023)}]{rmp_sensing_2}%
  \BibitemOpen
  \href@noop {} {\bibinfo {title} {{D.~F.~J. Kimball, D. Budker, T.~E. Chupp,
  A.~A. Geraci, S. Kolkowitz, J.~T. Singh, and A.~O. Sushkov, \textit{ Probing
  fundamental physics with spin-based quantum sensors}}}} (\bibinfo {year}
  {\href{https://journals.aps.org/pra/abstract/10.1103/PhysRevA.108.010101}{\textcolor{blue}{Rev.
  Mod. Phys. \textbf{108}, 010101 (2023)}}})\BibitemShut {NoStop}%
\bibitem [{cha(2004)}]{chang2004controlling}%
  \BibitemOpen
  \href@noop {} {\bibinfo {title} {{D.~E. Chang, J. Ye, and M.~D. Lukin,
  \textit{Controlling dipole-dipole frequency shifts in a lattice-based optical
  atomic clock}}}} (\bibinfo {year}
  {\href{https://journals.aps.org/pra/abstract/10.1103/PhysRevA.69.023810}{\textcolor{blue}{Phys.
  Rev. A \textbf{69}, 023810 (2004)}}})\BibitemShut {NoStop}%
\bibitem [{hut(2024)}]{hutson2023observation}%
  \BibitemOpen
  \href@noop {} {\bibinfo {title} {{R.~B. Hutson, W.~R. Milner, L. Yan, J. Ye,
  C. Sanner, \textit{Observation of millihertz-level cooperative Lamb shifts in
  an optical atomic clock}}}} (\bibinfo {year}
  {\href{https://www.science.org/doi/full/10.1126/science.adh4477}{\textcolor{blue}{Science
  \textbf{383}, 384 (2024)}}})\BibitemShut {NoStop}%
\bibitem [{bot(2022)}]{bothwell2022resolving}%
  \BibitemOpen
  \href@noop {} {\bibinfo {title} {{T. Bothwell, C.~J. Kennedy, A. Aeppli, D.
  Kedar, J.~M. Robinson, E. Oelker, A. Staron, and J. Ye, \textit{Resolving the
  gravitational redshift across a millimetre-scale atomic sample}}}} (\bibinfo
  {year}
  {\href{https://www.nature.com/articles/s41586-021-04349-7}{\textcolor{blue}{Nature
  (London) \textbf{602}, 420 (2022)}}})\BibitemShut {NoStop}%
\bibitem [{\citenamefont {Robinson}\ \emph {et~al.}(2024)\citenamefont
  {Robinson}, \citenamefont {Miklos}, \citenamefont {Tso}, \citenamefont
  {Kennedy}, \citenamefont {Kedar}, \citenamefont {Thompson},\ and\
  \citenamefont {Ye}}]{Robinson2023entanglement}%
  \BibitemOpen
  \bibfield  {author} {\bibinfo {author} {\bibfnamefont {J.~M.}\ \bibnamefont
  {Robinson}}, \bibinfo {author} {\bibfnamefont {M.}~\bibnamefont {Miklos}},
  \bibinfo {author} {\bibfnamefont {Y.~M.}\ \bibnamefont {Tso}}, \bibinfo
  {author} {\bibfnamefont {C.~J.}\ \bibnamefont {Kennedy}}, \bibinfo {author}
  {\bibfnamefont {D.}~\bibnamefont {Kedar}}, \bibinfo {author} {\bibfnamefont
  {J.~K.}\ \bibnamefont {Thompson}},\ and\ \bibinfo {author} {\bibfnamefont
  {J.}~\bibnamefont {Ye}},\ }\bibfield  {title} {\bibinfo {title} {Direct
  comparison of two spin squeezed optical clocks below the quantum projection
  noise limit},\ }\href {https://arxiv.org/abs/2211.08621} {\bibfield
  {journal} {\bibinfo  {journal} {Nature Physics}\ }\textbf {\bibinfo {volume}
  {20}},\ \bibinfo {pages} {208} (\bibinfo {year} {2024})}\BibitemShut
  {NoStop}%
\bibitem [{\citenamefont {Eckner}\ \emph {et~al.}(2023)\citenamefont {Eckner},
  \citenamefont {Darkwah~Oppong}, \citenamefont {Cao}, \citenamefont {Young},
  \citenamefont {Milner}, \citenamefont {Robinson}, \citenamefont {Ye},\ and\
  \citenamefont {Kaufman}}]{eckner2023realizing}%
  \BibitemOpen
  \bibfield  {author} {\bibinfo {author} {\bibfnamefont {W.~J.}\ \bibnamefont
  {Eckner}}, \bibinfo {author} {\bibfnamefont {N.}~\bibnamefont
  {Darkwah~Oppong}}, \bibinfo {author} {\bibfnamefont {A.}~\bibnamefont {Cao}},
  \bibinfo {author} {\bibfnamefont {A.~W.}\ \bibnamefont {Young}}, \bibinfo
  {author} {\bibfnamefont {W.~R.}\ \bibnamefont {Milner}}, \bibinfo {author}
  {\bibfnamefont {J.~M.}\ \bibnamefont {Robinson}}, \bibinfo {author}
  {\bibfnamefont {J.}~\bibnamefont {Ye}},\ and\ \bibinfo {author}
  {\bibfnamefont {A.~M.}\ \bibnamefont {Kaufman}},\ }\bibfield  {title}
  {\bibinfo {title} {Realizing spin squeezing with {Rydberg} interactions in an
  optical clock},\ }\href {https://www.nature.com/articles/s41586-023-06360-6}
  {\bibfield  {journal} {\bibinfo  {journal} {Nature (London)}\ }\textbf
  {\bibinfo {volume} {621}},\ \bibinfo {pages} {734} (\bibinfo {year}
  {2023})}\BibitemShut {NoStop}%
\bibitem [{\citenamefont {Lemonde}\ and\ \citenamefont
  {Wolf}(2005)}]{lemonde2005optical}%
  \BibitemOpen
  \bibfield  {author} {\bibinfo {author} {\bibfnamefont {P.}~\bibnamefont
  {Lemonde}}\ and\ \bibinfo {author} {\bibfnamefont {P.}~\bibnamefont {Wolf}},\
  }\bibfield  {title} {\bibinfo {title} {Optical lattice clock with atoms
  confined in a shallow trap},\ }\href
  {https://journals.aps.org/pra/abstract/10.1103/PhysRevA.72.033409} {\bibfield
   {journal} {\bibinfo  {journal} {Phys. Rev. A}\ }\textbf {\bibinfo {volume}
  {72}},\ \bibinfo {pages} {033409} (\bibinfo {year} {2005})}\BibitemShut
  {NoStop}%
\bibitem [{\citenamefont {Kolkowitz}\ \emph {et~al.}(2017)\citenamefont
  {Kolkowitz}, \citenamefont {Bromley}, \citenamefont {Bothwell}, \citenamefont
  {Wall}, \citenamefont {Marti}, \citenamefont {Koller}, \citenamefont {Zhang},
  \citenamefont {Rey},\ and\ \citenamefont {Ye}}]{kolkowitz2017spin}%
  \BibitemOpen
  \bibfield  {author} {\bibinfo {author} {\bibfnamefont {S.}~\bibnamefont
  {Kolkowitz}}, \bibinfo {author} {\bibfnamefont {S.}~\bibnamefont {Bromley}},
  \bibinfo {author} {\bibfnamefont {T.}~\bibnamefont {Bothwell}}, \bibinfo
  {author} {\bibfnamefont {M.}~\bibnamefont {Wall}}, \bibinfo {author}
  {\bibfnamefont {G.}~\bibnamefont {Marti}}, \bibinfo {author} {\bibfnamefont
  {A.}~\bibnamefont {Koller}}, \bibinfo {author} {\bibfnamefont
  {X.}~\bibnamefont {Zhang}}, \bibinfo {author} {\bibfnamefont
  {A.}~\bibnamefont {Rey}},\ and\ \bibinfo {author} {\bibfnamefont
  {J.}~\bibnamefont {Ye}},\ }\bibfield  {title} {\bibinfo {title}
  {Spin--orbit-coupled fermions in an optical lattice clock},\ }\href
  {https://www.nature.com/articles/nature20811} {\bibfield  {journal} {\bibinfo
   {journal} {Nature}\ }\textbf {\bibinfo {volume} {542}},\ \bibinfo {pages}
  {66} (\bibinfo {year} {2017})}\BibitemShut {NoStop}%
\bibitem [{\citenamefont {D{\"o}rscher}\ \emph {et~al.}(2018)\citenamefont
  {D{\"o}rscher}, \citenamefont {Schwarz}, \citenamefont {Al-Masoudi},
  \citenamefont {Falke}, \citenamefont {Sterr},\ and\ \citenamefont
  {Lisdat}}]{dorscher2018lattice}%
  \BibitemOpen
  \bibfield  {author} {\bibinfo {author} {\bibfnamefont {S.}~\bibnamefont
  {D{\"o}rscher}}, \bibinfo {author} {\bibfnamefont {R.}~\bibnamefont
  {Schwarz}}, \bibinfo {author} {\bibfnamefont {A.}~\bibnamefont {Al-Masoudi}},
  \bibinfo {author} {\bibfnamefont {S.}~\bibnamefont {Falke}}, \bibinfo
  {author} {\bibfnamefont {U.}~\bibnamefont {Sterr}},\ and\ \bibinfo {author}
  {\bibfnamefont {C.}~\bibnamefont {Lisdat}},\ }\bibfield  {title} {\bibinfo
  {title} {Lattice-induced photon scattering in an optical lattice clock},\
  }\href {https://journals.aps.org/pra/abstract/10.1103/PhysRevA.97.063419}
  {\bibfield  {journal} {\bibinfo  {journal} {Phys. Rev. A}\ }\textbf {\bibinfo
  {volume} {97}},\ \bibinfo {pages} {063419} (\bibinfo {year}
  {2018})}\BibitemShut {NoStop}%
\bibitem [{\citenamefont {Hutson}\ \emph {et~al.}(2019)\citenamefont {Hutson},
  \citenamefont {Goban}, \citenamefont {Marti}, \citenamefont {Sonderhouse},
  \citenamefont {Sanner},\ and\ \citenamefont {Ye}}]{hutson2019engineering}%
  \BibitemOpen
  \bibfield  {author} {\bibinfo {author} {\bibfnamefont {R.~B.}\ \bibnamefont
  {Hutson}}, \bibinfo {author} {\bibfnamefont {A.}~\bibnamefont {Goban}},
  \bibinfo {author} {\bibfnamefont {G.~E.}\ \bibnamefont {Marti}}, \bibinfo
  {author} {\bibfnamefont {L.}~\bibnamefont {Sonderhouse}}, \bibinfo {author}
  {\bibfnamefont {C.}~\bibnamefont {Sanner}},\ and\ \bibinfo {author}
  {\bibfnamefont {J.}~\bibnamefont {Ye}},\ }\bibfield  {title} {\bibinfo
  {title} {Engineering quantum states of matter for atomic clocks in shallow
  optical lattices},\ }\href
  {https://journals.aps.org/prl/abstract/10.1103/PhysRevLett.123.123401}
  {\bibfield  {journal} {\bibinfo  {journal} {Phys. Rev. Lett.}\ }\textbf
  {\bibinfo {volume} {123}},\ \bibinfo {pages} {123401} (\bibinfo {year}
  {2019})}\BibitemShut {NoStop}%
\bibitem [{\citenamefont {Duan}\ \emph {et~al.}(2003)\citenamefont {Duan},
  \citenamefont {Demler},\ and\ \citenamefont {Lukin}}]{duan2003controlling}%
  \BibitemOpen
  \bibfield  {author} {\bibinfo {author} {\bibfnamefont {L.-M.}\ \bibnamefont
  {Duan}}, \bibinfo {author} {\bibfnamefont {E.}~\bibnamefont {Demler}},\ and\
  \bibinfo {author} {\bibfnamefont {M.~D.}\ \bibnamefont {Lukin}},\ }\bibfield
  {title} {\bibinfo {title} {Controlling spin exchange interactions of
  ultracold atoms in optical lattices},\ }\href
  {https://journals.aps.org/prl/abstract/10.1103/PhysRevLett.91.090402}
  {\bibfield  {journal} {\bibinfo  {journal} {Phys. Rev. Lett.}\ }\textbf
  {\bibinfo {volume} {91}},\ \bibinfo {pages} {090402} (\bibinfo {year}
  {2003})}\BibitemShut {NoStop}%
\bibitem [{\citenamefont {Trotzky}\ \emph {et~al.}(2008)\citenamefont
  {Trotzky}, \citenamefont {Cheinet}, \citenamefont {Folling}, \citenamefont
  {Feld}, \citenamefont {Schnorrberger}, \citenamefont {Rey}, \citenamefont
  {Polkovnikov}, \citenamefont {Demler}, \citenamefont {Lukin},\ and\
  \citenamefont {Bloch}}]{trotzky2008time}%
  \BibitemOpen
  \bibfield  {author} {\bibinfo {author} {\bibfnamefont {S.}~\bibnamefont
  {Trotzky}}, \bibinfo {author} {\bibfnamefont {P.}~\bibnamefont {Cheinet}},
  \bibinfo {author} {\bibfnamefont {S.}~\bibnamefont {Folling}}, \bibinfo
  {author} {\bibfnamefont {M.}~\bibnamefont {Feld}}, \bibinfo {author}
  {\bibfnamefont {U.}~\bibnamefont {Schnorrberger}}, \bibinfo {author}
  {\bibfnamefont {A.~M.}\ \bibnamefont {Rey}}, \bibinfo {author} {\bibfnamefont
  {A.}~\bibnamefont {Polkovnikov}}, \bibinfo {author} {\bibfnamefont {E.~A.}\
  \bibnamefont {Demler}}, \bibinfo {author} {\bibfnamefont {M.~D.}\
  \bibnamefont {Lukin}},\ and\ \bibinfo {author} {\bibfnamefont
  {I.}~\bibnamefont {Bloch}},\ }\bibfield  {title} {\bibinfo {title}
  {Time-resolved observation and control of superexchange interactions with
  ultracold atoms in optical lattices},\ }\href
  {https://www.science.org/doi/10.1126/science.1150841} {\bibfield  {journal}
  {\bibinfo  {journal} {Science}\ }\textbf {\bibinfo {volume} {319}},\ \bibinfo
  {pages} {295} (\bibinfo {year} {2008})}\BibitemShut {NoStop}%
\bibitem [{\citenamefont {Celi}\ \emph {et~al.}(2014)\citenamefont {Celi},
  \citenamefont {Massignan}, \citenamefont {Ruseckas}, \citenamefont {Goldman},
  \citenamefont {Spielman}, \citenamefont {Juzeli{\=u}nas},\ and\ \citenamefont
  {Lewenstein}}]{celi2014synthetic}%
  \BibitemOpen
  \bibfield  {author} {\bibinfo {author} {\bibfnamefont {A.}~\bibnamefont
  {Celi}}, \bibinfo {author} {\bibfnamefont {P.}~\bibnamefont {Massignan}},
  \bibinfo {author} {\bibfnamefont {J.}~\bibnamefont {Ruseckas}}, \bibinfo
  {author} {\bibfnamefont {N.}~\bibnamefont {Goldman}}, \bibinfo {author}
  {\bibfnamefont {I.~B.}\ \bibnamefont {Spielman}}, \bibinfo {author}
  {\bibfnamefont {G.}~\bibnamefont {Juzeli{\=u}nas}},\ and\ \bibinfo {author}
  {\bibfnamefont {M.}~\bibnamefont {Lewenstein}},\ }\bibfield  {title}
  {\bibinfo {title} {Synthetic gauge fields in synthetic dimensions},\ }\href
  {https://journals.aps.org/prl/abstract/10.1103/PhysRevLett.112.043001}
  {\bibfield  {journal} {\bibinfo  {journal} {Phys. Rev. Lett.}\ }\textbf
  {\bibinfo {volume} {112}},\ \bibinfo {pages} {043001} (\bibinfo {year}
  {2014})}\BibitemShut {NoStop}%
\bibitem [{\citenamefont {Barbarino}\ \emph {et~al.}(2016)\citenamefont
  {Barbarino}, \citenamefont {Taddia}, \citenamefont {Rossini}, \citenamefont
  {Mazza},\ and\ \citenamefont {Fazio}}]{barbarino2016synthetic}%
  \BibitemOpen
  \bibfield  {author} {\bibinfo {author} {\bibfnamefont {S.}~\bibnamefont
  {Barbarino}}, \bibinfo {author} {\bibfnamefont {L.}~\bibnamefont {Taddia}},
  \bibinfo {author} {\bibfnamefont {D.}~\bibnamefont {Rossini}}, \bibinfo
  {author} {\bibfnamefont {L.}~\bibnamefont {Mazza}},\ and\ \bibinfo {author}
  {\bibfnamefont {R.}~\bibnamefont {Fazio}},\ }\bibfield  {title} {\bibinfo
  {title} {Synthetic gauge fields in synthetic dimensions: interactions and
  chiral edge modes},\ }\href
  {https://iopscience.iop.org/article/10.1088/1367-2630/18/3/035010} {\bibfield
   {journal} {\bibinfo  {journal} {New Journal of Physics}\ }\textbf {\bibinfo
  {volume} {18}},\ \bibinfo {pages} {035010} (\bibinfo {year}
  {2016})}\BibitemShut {NoStop}%
\bibitem [{\citenamefont {Aidelsburger}\ \emph {et~al.}(2013)\citenamefont
  {Aidelsburger}, \citenamefont {Atala}, \citenamefont {Lohse}, \citenamefont
  {Barreiro}, \citenamefont {Paredes},\ and\ \citenamefont
  {Bloch}}]{aidelsburger2013realization}%
  \BibitemOpen
  \bibfield  {author} {\bibinfo {author} {\bibfnamefont {M.}~\bibnamefont
  {Aidelsburger}}, \bibinfo {author} {\bibfnamefont {M.}~\bibnamefont {Atala}},
  \bibinfo {author} {\bibfnamefont {M.}~\bibnamefont {Lohse}}, \bibinfo
  {author} {\bibfnamefont {J.~T.}\ \bibnamefont {Barreiro}}, \bibinfo {author}
  {\bibfnamefont {B.}~\bibnamefont {Paredes}},\ and\ \bibinfo {author}
  {\bibfnamefont {I.}~\bibnamefont {Bloch}},\ }\bibfield  {title} {\bibinfo
  {title} {Realization of the hofstadter hamiltonian with ultracold atoms in
  optical lattices},\ }\href
  {https://journals.aps.org/prl/abstract/10.1103/PhysRevLett.111.185301}
  {\bibfield  {journal} {\bibinfo  {journal} {Phys. Rev. Lett.}\ }\textbf
  {\bibinfo {volume} {111}},\ \bibinfo {pages} {185301} (\bibinfo {year}
  {2013})}\BibitemShut {NoStop}%
\bibitem [{\citenamefont {Livi}\ \emph {et~al.}(2016)\citenamefont {Livi},
  \citenamefont {Cappellini}, \citenamefont {Diem}, \citenamefont {Franchi},
  \citenamefont {Clivati}, \citenamefont {Frittelli}, \citenamefont {Levi},
  \citenamefont {Calonico}, \citenamefont {Catani}, \citenamefont {Inguscio}
  \emph {et~al.}}]{livi2016synthetic}%
  \BibitemOpen
  \bibfield  {author} {\bibinfo {author} {\bibfnamefont {L.}~\bibnamefont
  {Livi}}, \bibinfo {author} {\bibfnamefont {G.}~\bibnamefont {Cappellini}},
  \bibinfo {author} {\bibfnamefont {M.}~\bibnamefont {Diem}}, \bibinfo {author}
  {\bibfnamefont {L.}~\bibnamefont {Franchi}}, \bibinfo {author} {\bibfnamefont
  {C.}~\bibnamefont {Clivati}}, \bibinfo {author} {\bibfnamefont
  {M.}~\bibnamefont {Frittelli}}, \bibinfo {author} {\bibfnamefont
  {F.}~\bibnamefont {Levi}}, \bibinfo {author} {\bibfnamefont {D.}~\bibnamefont
  {Calonico}}, \bibinfo {author} {\bibfnamefont {J.}~\bibnamefont {Catani}},
  \bibinfo {author} {\bibfnamefont {M.}~\bibnamefont {Inguscio}}, \emph
  {et~al.},\ }\bibfield  {title} {\bibinfo {title} {Synthetic dimensions and
  spin-orbit coupling with an optical clock transition},\ }\href
  {https://journals.aps.org/prl/abstract/10.1103/PhysRevLett.117.220401}
  {\bibfield  {journal} {\bibinfo  {journal} {Phys. Rev. Lett.}\ }\textbf
  {\bibinfo {volume} {117}},\ \bibinfo {pages} {220401} (\bibinfo {year}
  {2016})}\BibitemShut {NoStop}%
\bibitem [{\citenamefont {Tai}\ \emph {et~al.}(2017)\citenamefont {Tai},
  \citenamefont {Lukin}, \citenamefont {Rispoli}, \citenamefont {Schittko},
  \citenamefont {Menke}, \citenamefont {Borgnia}, \citenamefont {Preiss},
  \citenamefont {Grusdt}, \citenamefont {Kaufman},\ and\ \citenamefont
  {Greiner}}]{tai2017microscopy}%
  \BibitemOpen
  \bibfield  {author} {\bibinfo {author} {\bibfnamefont {M.~E.}\ \bibnamefont
  {Tai}}, \bibinfo {author} {\bibfnamefont {A.}~\bibnamefont {Lukin}}, \bibinfo
  {author} {\bibfnamefont {M.}~\bibnamefont {Rispoli}}, \bibinfo {author}
  {\bibfnamefont {R.}~\bibnamefont {Schittko}}, \bibinfo {author}
  {\bibfnamefont {T.}~\bibnamefont {Menke}}, \bibinfo {author} {\bibfnamefont
  {D.}~\bibnamefont {Borgnia}}, \bibinfo {author} {\bibfnamefont {P.~M.}\
  \bibnamefont {Preiss}}, \bibinfo {author} {\bibfnamefont {F.}~\bibnamefont
  {Grusdt}}, \bibinfo {author} {\bibfnamefont {A.~M.}\ \bibnamefont
  {Kaufman}},\ and\ \bibinfo {author} {\bibfnamefont {M.}~\bibnamefont
  {Greiner}},\ }\bibfield  {title} {\bibinfo {title} {Microscopy of the
  interacting harper--hofstadter model in the two-body limit},\ }\href
  {https://www.nature.com/articles/nature22811} {\bibfield  {journal} {\bibinfo
   {journal} {Nature}\ }\textbf {\bibinfo {volume} {546}},\ \bibinfo {pages}
  {519} (\bibinfo {year} {2017})}\BibitemShut {NoStop}%
\bibitem [{\citenamefont {Hild}\ \emph {et~al.}(2014)\citenamefont {Hild},
  \citenamefont {Fukuhara}, \citenamefont {Schau{\ss}}, \citenamefont {Zeiher},
  \citenamefont {Knap}, \citenamefont {Demler}, \citenamefont {Bloch},\ and\
  \citenamefont {Gross}}]{hild2014far}%
  \BibitemOpen
  \bibfield  {author} {\bibinfo {author} {\bibfnamefont {S.}~\bibnamefont
  {Hild}}, \bibinfo {author} {\bibfnamefont {T.}~\bibnamefont {Fukuhara}},
  \bibinfo {author} {\bibfnamefont {P.}~\bibnamefont {Schau{\ss}}}, \bibinfo
  {author} {\bibfnamefont {J.}~\bibnamefont {Zeiher}}, \bibinfo {author}
  {\bibfnamefont {M.}~\bibnamefont {Knap}}, \bibinfo {author} {\bibfnamefont
  {E.}~\bibnamefont {Demler}}, \bibinfo {author} {\bibfnamefont
  {I.}~\bibnamefont {Bloch}},\ and\ \bibinfo {author} {\bibfnamefont
  {C.}~\bibnamefont {Gross}},\ }\bibfield  {title} {\bibinfo {title}
  {Far-from-equilibrium spin transport in heisenberg quantum magnets},\ }\href
  {https://journals.aps.org/prl/abstract/10.1103/PhysRevLett.113.147205}
  {\bibfield  {journal} {\bibinfo  {journal} {Phys. Rev. Lett.}\ }\textbf
  {\bibinfo {volume} {113}},\ \bibinfo {pages} {147205} (\bibinfo {year}
  {2014})}\BibitemShut {NoStop}%
\bibitem [{\citenamefont {Jepsen}\ \emph {et~al.}(2021)\citenamefont {Jepsen},
  \citenamefont {Ho}, \citenamefont {Amato-Grill}, \citenamefont {Dimitrova},
  \citenamefont {Demler},\ and\ \citenamefont
  {Ketterle}}]{jepsen2021transverse}%
  \BibitemOpen
  \bibfield  {author} {\bibinfo {author} {\bibfnamefont {P.~N.}\ \bibnamefont
  {Jepsen}}, \bibinfo {author} {\bibfnamefont {W.~W.}\ \bibnamefont {Ho}},
  \bibinfo {author} {\bibfnamefont {J.}~\bibnamefont {Amato-Grill}}, \bibinfo
  {author} {\bibfnamefont {I.}~\bibnamefont {Dimitrova}}, \bibinfo {author}
  {\bibfnamefont {E.}~\bibnamefont {Demler}},\ and\ \bibinfo {author}
  {\bibfnamefont {W.}~\bibnamefont {Ketterle}},\ }\bibfield  {title} {\bibinfo
  {title} {Transverse spin dynamics in the anisotropic heisenberg model
  realized with ultracold atoms},\ }\href
  {https://journals.aps.org/prx/abstract/10.1103/PhysRevX.11.041054} {\bibfield
   {journal} {\bibinfo  {journal} {Phys. Rev. X}\ }\textbf {\bibinfo {volume}
  {11}},\ \bibinfo {pages} {041054} (\bibinfo {year} {2021})}\BibitemShut
  {NoStop}%
\bibitem [{\citenamefont {Jepsen}\ \emph {et~al.}(2022)\citenamefont {Jepsen},
  \citenamefont {Lee}, \citenamefont {Lin}, \citenamefont {Dimitrova},
  \citenamefont {Margalit}, \citenamefont {Ho},\ and\ \citenamefont
  {Ketterle}}]{jepsen2022long}%
  \BibitemOpen
  \bibfield  {author} {\bibinfo {author} {\bibfnamefont {P.~N.}\ \bibnamefont
  {Jepsen}}, \bibinfo {author} {\bibfnamefont {Y.~K.}\ \bibnamefont {Lee}},
  \bibinfo {author} {\bibfnamefont {H.}~\bibnamefont {Lin}}, \bibinfo {author}
  {\bibfnamefont {I.}~\bibnamefont {Dimitrova}}, \bibinfo {author}
  {\bibfnamefont {Y.}~\bibnamefont {Margalit}}, \bibinfo {author}
  {\bibfnamefont {W.~W.}\ \bibnamefont {Ho}},\ and\ \bibinfo {author}
  {\bibfnamefont {W.}~\bibnamefont {Ketterle}},\ }\bibfield  {title} {\bibinfo
  {title} {Long-lived phantom helix states in heisenberg quantum magnets},\
  }\href {https://www.nature.com/articles/s41567-022-01651-7} {\bibfield
  {journal} {\bibinfo  {journal} {Nature Physics}\ }\textbf {\bibinfo {volume}
  {18}},\ \bibinfo {pages} {899} (\bibinfo {year} {2022})}\BibitemShut
  {NoStop}%
\bibitem [{\citenamefont {Liang}\ \emph {et~al.}(2021)\citenamefont {Liang},
  \citenamefont {Trypogeorgos}, \citenamefont {Vald{\'e}s-Curiel},
  \citenamefont {Tao}, \citenamefont {Zhao},\ and\ \citenamefont
  {Spielman}}]{liang2021coherence}%
  \BibitemOpen
  \bibfield  {author} {\bibinfo {author} {\bibfnamefont {Q.-Y.}\ \bibnamefont
  {Liang}}, \bibinfo {author} {\bibfnamefont {D.}~\bibnamefont {Trypogeorgos}},
  \bibinfo {author} {\bibfnamefont {A.}~\bibnamefont {Vald{\'e}s-Curiel}},
  \bibinfo {author} {\bibfnamefont {J.}~\bibnamefont {Tao}}, \bibinfo {author}
  {\bibfnamefont {M.}~\bibnamefont {Zhao}},\ and\ \bibinfo {author}
  {\bibfnamefont {I.~B.}\ \bibnamefont {Spielman}},\ }\bibfield  {title}
  {\bibinfo {title} {Coherence and decoherence in the harper-hofstadter
  model},\ }\href
  {https://journals.aps.org/prresearch/abstract/10.1103/PhysRevResearch.3.023058}
  {\bibfield  {journal} {\bibinfo  {journal} {Phys. Rev. Res.}\ }\textbf
  {\bibinfo {volume} {3}},\ \bibinfo {pages} {023058} (\bibinfo {year}
  {2021})}\BibitemShut {NoStop}%
\bibitem [{\citenamefont {An}\ \emph {et~al.}(2017)\citenamefont {An},
  \citenamefont {Meier},\ and\ \citenamefont {Gadway}}]{an2017direct}%
  \BibitemOpen
  \bibfield  {author} {\bibinfo {author} {\bibfnamefont {F.~A.}\ \bibnamefont
  {An}}, \bibinfo {author} {\bibfnamefont {E.~J.}\ \bibnamefont {Meier}},\ and\
  \bibinfo {author} {\bibfnamefont {B.}~\bibnamefont {Gadway}},\ }\bibfield
  {title} {\bibinfo {title} {Direct observation of chiral currents and magnetic
  reflection in atomic flux lattices},\ }\href
  {https://www.science.org/doi/10.1126/sciadv.1602685} {\bibfield  {journal}
  {\bibinfo  {journal} {Science advances}\ }\textbf {\bibinfo {volume} {3}},\
  \bibinfo {pages} {e1602685} (\bibinfo {year} {2017})}\BibitemShut {NoStop}%
\bibitem [{\citenamefont {Marti}\ \emph {et~al.}(2018)\citenamefont {Marti},
  \citenamefont {Hutson}, \citenamefont {Goban}, \citenamefont {Campbell},
  \citenamefont {Poli},\ and\ \citenamefont {Ye}}]{marti2018imaging}%
  \BibitemOpen
  \bibfield  {author} {\bibinfo {author} {\bibfnamefont {G.~E.}\ \bibnamefont
  {Marti}}, \bibinfo {author} {\bibfnamefont {R.~B.}\ \bibnamefont {Hutson}},
  \bibinfo {author} {\bibfnamefont {A.}~\bibnamefont {Goban}}, \bibinfo
  {author} {\bibfnamefont {S.~L.}\ \bibnamefont {Campbell}}, \bibinfo {author}
  {\bibfnamefont {N.}~\bibnamefont {Poli}},\ and\ \bibinfo {author}
  {\bibfnamefont {J.}~\bibnamefont {Ye}},\ }\bibfield  {title} {\bibinfo
  {title} {Imaging optical frequencies with 100 $\mu$hz precision and 1.1
  $\mu$m resolution},\ }\href
  {https://journals.aps.org/prl/abstract/10.1103/PhysRevLett.120.103201}
  {\bibfield  {journal} {\bibinfo  {journal} {Phys. Rev. Lett.}\ }\textbf
  {\bibinfo {volume} {120}},\ \bibinfo {pages} {103201} (\bibinfo {year}
  {2018})}\BibitemShut {NoStop}%
\bibitem [{\citenamefont {Mamaev}\ \emph {et~al.}(2021)\citenamefont {Mamaev},
  \citenamefont {Kimchi}, \citenamefont {Nandkishore},\ and\ \citenamefont
  {Rey}}]{mamaev2021tunable}%
  \BibitemOpen
  \bibfield  {author} {\bibinfo {author} {\bibfnamefont {M.}~\bibnamefont
  {Mamaev}}, \bibinfo {author} {\bibfnamefont {I.}~\bibnamefont {Kimchi}},
  \bibinfo {author} {\bibfnamefont {R.~M.}\ \bibnamefont {Nandkishore}},\ and\
  \bibinfo {author} {\bibfnamefont {A.~M.}\ \bibnamefont {Rey}},\ }\bibfield
  {title} {\bibinfo {title} {Tunable-spin-model generation with
  spin-orbit-coupled fermions in optical lattices},\ }\href
  {https://journals.aps.org/prresearch/abstract/10.1103/PhysRevResearch.3.013178}
  {\bibfield  {journal} {\bibinfo  {journal} {Phys. Rev. Res.}\ }\textbf
  {\bibinfo {volume} {3}},\ \bibinfo {pages} {013178} (\bibinfo {year}
  {2021})}\BibitemShut {NoStop}%
\bibitem [{\citenamefont {Mamaev}\ \emph {et~al.}(2022)\citenamefont {Mamaev},
  \citenamefont {Bilitewski}, \citenamefont {Sundar},\ and\ \citenamefont
  {Rey}}]{mamaev2022resonant}%
  \BibitemOpen
  \bibfield  {author} {\bibinfo {author} {\bibfnamefont {M.}~\bibnamefont
  {Mamaev}}, \bibinfo {author} {\bibfnamefont {T.}~\bibnamefont {Bilitewski}},
  \bibinfo {author} {\bibfnamefont {B.}~\bibnamefont {Sundar}},\ and\ \bibinfo
  {author} {\bibfnamefont {A.~M.}\ \bibnamefont {Rey}},\ }\bibfield  {title}
  {\bibinfo {title} {Resonant dynamics of strongly interacting su(n) fermionic
  atoms in a synthetic flux ladder},\ }\href
  {https://journals.aps.org/prxquantum/abstract/10.1103/PRXQuantum.3.030328}
  {\bibfield  {journal} {\bibinfo  {journal} {PRX Quantum}\ }\textbf {\bibinfo
  {volume} {3}},\ \bibinfo {pages} {030328} (\bibinfo {year}
  {2022})}\BibitemShut {NoStop}%
\bibitem [{\citenamefont {Bukov}\ \emph {et~al.}(2016)\citenamefont {Bukov},
  \citenamefont {Kolodrubetz},\ and\ \citenamefont
  {Polkovnikov}}]{bukov2016schrieffer}%
  \BibitemOpen
  \bibfield  {author} {\bibinfo {author} {\bibfnamefont {M.}~\bibnamefont
  {Bukov}}, \bibinfo {author} {\bibfnamefont {M.}~\bibnamefont {Kolodrubetz}},\
  and\ \bibinfo {author} {\bibfnamefont {A.}~\bibnamefont {Polkovnikov}},\
  }\bibfield  {title} {\bibinfo {title} {Schrieffer-wolff transformation for
  periodically driven systems: Strongly correlated systems with artificial
  gauge fields},\ }\href
  {https://journals.aps.org/prl/abstract/10.1103/PhysRevLett.116.125301}
  {\bibfield  {journal} {\bibinfo  {journal} {Phys. Rev. Lett.}\ }\textbf
  {\bibinfo {volume} {116}},\ \bibinfo {pages} {125301} (\bibinfo {year}
  {2016})}\BibitemShut {NoStop}%
\bibitem [{\citenamefont {Mamaev}\ \emph {et~al.}(2019)\citenamefont {Mamaev},
  \citenamefont {Kimchi}, \citenamefont {Perlin}, \citenamefont {Nandkishore},\
  and\ \citenamefont {Rey}}]{mamaev2019quantum}%
  \BibitemOpen
  \bibfield  {author} {\bibinfo {author} {\bibfnamefont {M.}~\bibnamefont
  {Mamaev}}, \bibinfo {author} {\bibfnamefont {I.}~\bibnamefont {Kimchi}},
  \bibinfo {author} {\bibfnamefont {M.~A.}\ \bibnamefont {Perlin}}, \bibinfo
  {author} {\bibfnamefont {R.~M.}\ \bibnamefont {Nandkishore}},\ and\ \bibinfo
  {author} {\bibfnamefont {A.~M.}\ \bibnamefont {Rey}},\ }\bibfield  {title}
  {\bibinfo {title} {Quantum entropic self-localization with ultracold
  fermions},\ }\href
  {https://journals.aps.org/prl/abstract/10.1103/PhysRevLett.123.130402}
  {\bibfield  {journal} {\bibinfo  {journal} {Phys. Rev. Lett.}\ }\textbf
  {\bibinfo {volume} {123}},\ \bibinfo {pages} {130402} (\bibinfo {year}
  {2019})}\BibitemShut {NoStop}%
\bibitem [{\citenamefont {Xu}\ \emph {et~al.}(2018)\citenamefont {Xu},
  \citenamefont {Morong}, \citenamefont {Hui}, \citenamefont {Scarola},\ and\
  \citenamefont {DeMarco}}]{xu2018correlated}%
  \BibitemOpen
  \bibfield  {author} {\bibinfo {author} {\bibfnamefont {W.}~\bibnamefont
  {Xu}}, \bibinfo {author} {\bibfnamefont {W.}~\bibnamefont {Morong}}, \bibinfo
  {author} {\bibfnamefont {H.-Y.}\ \bibnamefont {Hui}}, \bibinfo {author}
  {\bibfnamefont {V.~W.}\ \bibnamefont {Scarola}},\ and\ \bibinfo {author}
  {\bibfnamefont {B.}~\bibnamefont {DeMarco}},\ }\bibfield  {title} {\bibinfo
  {title} {Correlated spin-flip tunneling in a fermi lattice gas},\ }\href
  {https://journals.aps.org/pra/abstract/10.1103/PhysRevA.98.023623} {\bibfield
   {journal} {\bibinfo  {journal} {Phys. Rev. A}\ }\textbf {\bibinfo {volume}
  {98}},\ \bibinfo {pages} {023623} (\bibinfo {year} {2018})}\BibitemShut
  {NoStop}%
\bibitem [{\citenamefont {Milner}\ \emph {et~al.}(2023)\citenamefont {Milner},
  \citenamefont {Yan}, \citenamefont {Hutson}, \citenamefont {Sanner},\ and\
  \citenamefont {Ye}}]{milner2023high}%
  \BibitemOpen
  \bibfield  {author} {\bibinfo {author} {\bibfnamefont {W.~R.}\ \bibnamefont
  {Milner}}, \bibinfo {author} {\bibfnamefont {L.}~\bibnamefont {Yan}},
  \bibinfo {author} {\bibfnamefont {R.~B.}\ \bibnamefont {Hutson}}, \bibinfo
  {author} {\bibfnamefont {C.}~\bibnamefont {Sanner}},\ and\ \bibinfo {author}
  {\bibfnamefont {J.}~\bibnamefont {Ye}},\ }\bibfield  {title} {\bibinfo
  {title} {High-fidelity imaging of a band insulator in a three-dimensional
  optical lattice clock},\ }\href
  {https://journals.aps.org/pra/abstract/10.1103/PhysRevA.107.063313}
  {\bibfield  {journal} {\bibinfo  {journal} {Phys. Rev. A}\ }\textbf {\bibinfo
  {volume} {107}},\ \bibinfo {pages} {063313} (\bibinfo {year}
  {2023})}\BibitemShut {NoStop}%
\bibitem [{\citenamefont {Mukherjee}\ \emph {et~al.}(2017)\citenamefont
  {Mukherjee}, \citenamefont {Yan}, \citenamefont {Patel}, \citenamefont
  {Hadzibabic}, \citenamefont {Yefsah}, \citenamefont {Struck},\ and\
  \citenamefont {Zwierlein}}]{mukherjee2017homogeneous}%
  \BibitemOpen
  \bibfield  {author} {\bibinfo {author} {\bibfnamefont {B.}~\bibnamefont
  {Mukherjee}}, \bibinfo {author} {\bibfnamefont {Z.}~\bibnamefont {Yan}},
  \bibinfo {author} {\bibfnamefont {P.~B.}\ \bibnamefont {Patel}}, \bibinfo
  {author} {\bibfnamefont {Z.}~\bibnamefont {Hadzibabic}}, \bibinfo {author}
  {\bibfnamefont {T.}~\bibnamefont {Yefsah}}, \bibinfo {author} {\bibfnamefont
  {J.}~\bibnamefont {Struck}},\ and\ \bibinfo {author} {\bibfnamefont {M.~W.}\
  \bibnamefont {Zwierlein}},\ }\bibfield  {title} {\bibinfo {title}
  {Homogeneous atomic fermi gases},\ }\href
  {https://journals.aps.org/prl/abstract/10.1103/PhysRevLett.118.123401}
  {\bibfield  {journal} {\bibinfo  {journal} {Phys. Rev. Lett.}\ }\textbf
  {\bibinfo {volume} {118}},\ \bibinfo {pages} {123401} (\bibinfo {year}
  {2017})}\BibitemShut {NoStop}%
\bibitem [{\citenamefont {Omran}\ \emph {et~al.}(2015)\citenamefont {Omran},
  \citenamefont {Boll}, \citenamefont {Hilker}, \citenamefont {Kleinlein},
  \citenamefont {Salomon}, \citenamefont {Bloch},\ and\ \citenamefont
  {Gross}}]{omran2015microscopic}%
  \BibitemOpen
  \bibfield  {author} {\bibinfo {author} {\bibfnamefont {A.}~\bibnamefont
  {Omran}}, \bibinfo {author} {\bibfnamefont {M.}~\bibnamefont {Boll}},
  \bibinfo {author} {\bibfnamefont {T.~A.}\ \bibnamefont {Hilker}}, \bibinfo
  {author} {\bibfnamefont {K.}~\bibnamefont {Kleinlein}}, \bibinfo {author}
  {\bibfnamefont {G.}~\bibnamefont {Salomon}}, \bibinfo {author} {\bibfnamefont
  {I.}~\bibnamefont {Bloch}},\ and\ \bibinfo {author} {\bibfnamefont
  {C.}~\bibnamefont {Gross}},\ }\bibfield  {title} {\bibinfo {title}
  {Microscopic observation of pauli blocking in degenerate fermionic lattice
  gases},\ }\href
  {https://journals.aps.org/prl/abstract/10.1103/PhysRevLett.115.263001}
  {\bibfield  {journal} {\bibinfo  {journal} {Phys. Rev. Lett.}\ }\textbf
  {\bibinfo {volume} {115}},\ \bibinfo {pages} {263001} (\bibinfo {year}
  {2015})}\BibitemShut {NoStop}%
\bibitem [{\citenamefont {Dimitrova}\ \emph {et~al.}(2020)\citenamefont
  {Dimitrova}, \citenamefont {Jepsen}, \citenamefont {Buyskikh}, \citenamefont
  {Venegas-Gomez}, \citenamefont {Amato-Grill}, \citenamefont {Daley},\ and\
  \citenamefont {Ketterle}}]{dimitrova2020enhanced}%
  \BibitemOpen
  \bibfield  {author} {\bibinfo {author} {\bibfnamefont {I.}~\bibnamefont
  {Dimitrova}}, \bibinfo {author} {\bibfnamefont {N.}~\bibnamefont {Jepsen}},
  \bibinfo {author} {\bibfnamefont {A.}~\bibnamefont {Buyskikh}}, \bibinfo
  {author} {\bibfnamefont {A.}~\bibnamefont {Venegas-Gomez}}, \bibinfo {author}
  {\bibfnamefont {J.}~\bibnamefont {Amato-Grill}}, \bibinfo {author}
  {\bibfnamefont {A.}~\bibnamefont {Daley}},\ and\ \bibinfo {author}
  {\bibfnamefont {W.}~\bibnamefont {Ketterle}},\ }\bibfield  {title} {\bibinfo
  {title} {Enhanced superexchange in a tilted mott insulator},\ }\href
  {https://journals.aps.org/prl/abstract/10.1103/PhysRevLett.124.043204}
  {\bibfield  {journal} {\bibinfo  {journal} {Phys. Rev. Lett.}\ }\textbf
  {\bibinfo {volume} {124}},\ \bibinfo {pages} {043204} (\bibinfo {year}
  {2020})}\BibitemShut {NoStop}%
\bibitem [{\citenamefont {Foss-Feig}\ \emph {et~al.}(2013)\citenamefont
  {Foss-Feig}, \citenamefont {Hazzard}, \citenamefont {Bollinger},\ and\
  \citenamefont {Rey}}]{foss2013nonequilibrium}%
  \BibitemOpen
  \bibfield  {author} {\bibinfo {author} {\bibfnamefont {M.}~\bibnamefont
  {Foss-Feig}}, \bibinfo {author} {\bibfnamefont {K.~R.}\ \bibnamefont
  {Hazzard}}, \bibinfo {author} {\bibfnamefont {J.~J.}\ \bibnamefont
  {Bollinger}},\ and\ \bibinfo {author} {\bibfnamefont {A.~M.}\ \bibnamefont
  {Rey}},\ }\bibfield  {title} {\bibinfo {title} {Nonequilibrium dynamics of
  arbitrary-range ising models with decoherence: An exact analytic solution},\
  }\href {https://journals.aps.org/pra/abstract/10.1103/PhysRevA.87.042101}
  {\bibfield  {journal} {\bibinfo  {journal} {Phys. Rev. A}\ }\textbf {\bibinfo
  {volume} {87}},\ \bibinfo {pages} {042101} (\bibinfo {year}
  {2013})}\BibitemShut {NoStop}%
\bibitem [{\citenamefont {Schachenmayer}\ \emph {et~al.}(2015)\citenamefont
  {Schachenmayer}, \citenamefont {Pikovski},\ and\ \citenamefont
  {Rey}}]{schachenmayer2015many}%
  \BibitemOpen
  \bibfield  {author} {\bibinfo {author} {\bibfnamefont {J.}~\bibnamefont
  {Schachenmayer}}, \bibinfo {author} {\bibfnamefont {A.}~\bibnamefont
  {Pikovski}},\ and\ \bibinfo {author} {\bibfnamefont {A.~M.}\ \bibnamefont
  {Rey}},\ }\bibfield  {title} {\bibinfo {title} {Many-body quantum spin
  dynamics with monte carlo trajectories on a discrete phase space},\ }\href
  {https://journals.aps.org/prx/abstract/10.1103/PhysRevX.5.011022} {\bibfield
  {journal} {\bibinfo  {journal} {Phys. Rev. X}\ }\textbf {\bibinfo {volume}
  {5}},\ \bibinfo {pages} {011022} (\bibinfo {year} {2015})}\BibitemShut
  {NoStop}%
\bibitem [{\citenamefont {Kim}\ \emph {et~al.}(2025)\citenamefont {Kim},
  \citenamefont {Aeppli}, \citenamefont {Warfield}, \citenamefont {Chu},
  \citenamefont {Rey},\ and\ \citenamefont {Ye}}]{kim2025atomic}%
  \BibitemOpen
  \bibfield  {author} {\bibinfo {author} {\bibfnamefont {K.}~\bibnamefont
  {Kim}}, \bibinfo {author} {\bibfnamefont {A.}~\bibnamefont {Aeppli}},
  \bibinfo {author} {\bibfnamefont {W.}~\bibnamefont {Warfield}}, \bibinfo
  {author} {\bibfnamefont {A.}~\bibnamefont {Chu}}, \bibinfo {author}
  {\bibfnamefont {A.~M.}\ \bibnamefont {Rey}},\ and\ \bibinfo {author}
  {\bibfnamefont {J.}~\bibnamefont {Ye}},\ }\bibfield  {title} {\bibinfo
  {title} {Atomic coherence of 2 minutes and instability of 1.5 e-18 at 1 s in
  a wannier-stark lattice clock},\ }\href {https://arxiv.org/pdf/2505.06444}
  {\bibfield  {journal} {\bibinfo  {journal} {arXiv preprint arXiv:2505.06444}\
  } (\bibinfo {year} {2025})}\BibitemShut {NoStop}%
\bibitem [{\citenamefont {Kitagawa}\ and\ \citenamefont
  {Ueda}(1993)}]{kitagawa1993squeezed}%
  \BibitemOpen
  \bibfield  {author} {\bibinfo {author} {\bibfnamefont {M.}~\bibnamefont
  {Kitagawa}}\ and\ \bibinfo {author} {\bibfnamefont {M.}~\bibnamefont
  {Ueda}},\ }\bibfield  {title} {\bibinfo {title} {Squeezed spin states},\
  }\href {https://journals.aps.org/pra/abstract/10.1103/PhysRevA.47.5138}
  {\bibfield  {journal} {\bibinfo  {journal} {Phys. Rev. A}\ }\textbf {\bibinfo
  {volume} {47}},\ \bibinfo {pages} {5138} (\bibinfo {year}
  {1993})}\BibitemShut {NoStop}%
\bibitem [{\citenamefont {Ma}\ \emph {et~al.}(2011)\citenamefont {Ma},
  \citenamefont {Wang}, \citenamefont {Sun},\ and\ \citenamefont
  {Nori}}]{ma2011quantum}%
  \BibitemOpen
  \bibfield  {author} {\bibinfo {author} {\bibfnamefont {J.}~\bibnamefont
  {Ma}}, \bibinfo {author} {\bibfnamefont {X.}~\bibnamefont {Wang}}, \bibinfo
  {author} {\bibfnamefont {C.-P.}\ \bibnamefont {Sun}},\ and\ \bibinfo {author}
  {\bibfnamefont {F.}~\bibnamefont {Nori}},\ }\bibfield  {title} {\bibinfo
  {title} {Quantum spin squeezing},\ }\href
  {https://www.sciencedirect.com/science/article/abs/pii/S0370157311002201}
  {\bibfield  {journal} {\bibinfo  {journal} {Physics Reports}\ }\textbf
  {\bibinfo {volume} {509}},\ \bibinfo {pages} {89} (\bibinfo {year}
  {2011})}\BibitemShut {NoStop}%
\bibitem [{\citenamefont {He}\ \emph {et~al.}(2019)\citenamefont {He},
  \citenamefont {Perlin}, \citenamefont {Muleady}, \citenamefont {Lewis-Swan},
  \citenamefont {Hutson}, \citenamefont {Ye},\ and\ \citenamefont
  {Rey}}]{he2019engineering}%
  \BibitemOpen
  \bibfield  {author} {\bibinfo {author} {\bibfnamefont {P.}~\bibnamefont
  {He}}, \bibinfo {author} {\bibfnamefont {M.~A.}\ \bibnamefont {Perlin}},
  \bibinfo {author} {\bibfnamefont {S.~R.}\ \bibnamefont {Muleady}}, \bibinfo
  {author} {\bibfnamefont {R.~J.}\ \bibnamefont {Lewis-Swan}}, \bibinfo
  {author} {\bibfnamefont {R.~B.}\ \bibnamefont {Hutson}}, \bibinfo {author}
  {\bibfnamefont {J.}~\bibnamefont {Ye}},\ and\ \bibinfo {author}
  {\bibfnamefont {A.~M.}\ \bibnamefont {Rey}},\ }\bibfield  {title} {\bibinfo
  {title} {Engineering spin squeezing in a 3d optical lattice with interacting
  spin-orbit-coupled fermions},\ }\href
  {https://journals.aps.org/prresearch/abstract/10.1103/PhysRevResearch.1.033075}
  {\bibfield  {journal} {\bibinfo  {journal} {Phys. Rev. Res.}\ }\textbf
  {\bibinfo {volume} {1}},\ \bibinfo {pages} {033075} (\bibinfo {year}
  {2019})}\BibitemShut {NoStop}%
\bibitem [{\citenamefont {Hern{\'a}ndez~Yanes}\ \emph
  {et~al.}(2022)\citenamefont {Hern{\'a}ndez~Yanes}, \citenamefont
  {P{\l}odzie{\'n}}, \citenamefont {Mackoit~Sinkevi{\v{c}}ien{\.e}},
  \citenamefont {{\v{Z}}labys}, \citenamefont {Juzeli{\=u}nas},\ and\
  \citenamefont {Witkowska}}]{hernandez2022one}%
  \BibitemOpen
  \bibfield  {author} {\bibinfo {author} {\bibfnamefont {T.}~\bibnamefont
  {Hern{\'a}ndez~Yanes}}, \bibinfo {author} {\bibfnamefont {M.}~\bibnamefont
  {P{\l}odzie{\'n}}}, \bibinfo {author} {\bibfnamefont {M.}~\bibnamefont
  {Mackoit~Sinkevi{\v{c}}ien{\.e}}}, \bibinfo {author} {\bibfnamefont
  {G.}~\bibnamefont {{\v{Z}}labys}}, \bibinfo {author} {\bibfnamefont
  {G.}~\bibnamefont {Juzeli{\=u}nas}},\ and\ \bibinfo {author} {\bibfnamefont
  {E.}~\bibnamefont {Witkowska}},\ }\bibfield  {title} {\bibinfo {title}
  {One-and two-axis squeezing via laser coupling in an atomic fermi-hubbard
  model},\ }\href
  {https://journals.aps.org/prl/abstract/10.1103/PhysRevLett.129.090403}
  {\bibfield  {journal} {\bibinfo  {journal} {Phys. Rev. Lett.}\ }\textbf
  {\bibinfo {volume} {129}},\ \bibinfo {pages} {090403} (\bibinfo {year}
  {2022})}\BibitemShut {NoStop}%
\bibitem [{\citenamefont {Mamaev}\ \emph {et~al.}(2024)\citenamefont {Mamaev},
  \citenamefont {Barberena},\ and\ \citenamefont {Rey}}]{mamaev2024spin}%
  \BibitemOpen
  \bibfield  {author} {\bibinfo {author} {\bibfnamefont {M.}~\bibnamefont
  {Mamaev}}, \bibinfo {author} {\bibfnamefont {D.}~\bibnamefont {Barberena}},\
  and\ \bibinfo {author} {\bibfnamefont {A.~M.}\ \bibnamefont {Rey}},\
  }\bibfield  {title} {\bibinfo {title} {Spin squeezing in mixed-dimensional
  anisotropic lattice models},\ }\href
  {https://journals.aps.org/pra/abstract/10.1103/PhysRevA.109.023326}
  {\bibfield  {journal} {\bibinfo  {journal} {Phys. Rev. A}\ }\textbf {\bibinfo
  {volume} {109}},\ \bibinfo {pages} {023326} (\bibinfo {year}
  {2024})}\BibitemShut {NoStop}%
\end{thebibliography}%

\clearpage
\onecolumngrid
\appendix

%%%%%
\section{Superexchange and $t-J$ model derivations}
\label{app_Superexchange}
%%%%%

In the main text, we describe an effective superexchange spin-1/2 model to capture the physics of the full Fermi-Hubbard model, which we re-write here (in the gauged frame where the clock laser phase has been put into the tunneling):
\begin{equation}
\hat{H}' = - t_z \sum_{j}\left(e^{i \phi/2}\hat{c}_{j,e}^{\dagger}\hat{c}_{j+1,e} +e^{-i \phi/2}\hat{c}_{j,g}^{\dagger}\hat{c}_{j+1,g}+ h.c.\right)+ U \sum_{j}\hat{n}_{j,e}\hat{n}_{j,g}+ \eta_z\sum_{j} (j-j_0)^2 \left(\hat{n}_{j,e} + \hat{n}_{j,g}\right).
\end{equation}
We omit the Rabi drive for now, and will add it at the end of the derivation.

The spin-1/2 model can be derived with standard Schrieffer-Wolff second order perturbation theory. This has been worked out in Refs.~\cite{mamaev2021tunable,mamaev2024spin}, though we spell it out here again, since it will be instructive for the more complex models derived further on. We consider two neighbouring sites $j$, $j+1$ filled by two atoms. The Hilbert space contains six states $\{\ket{e,e}, \ket{e,g}, \ket{g,e},\ket{g,g}, \ket{eg,0}, \ket{0,eg}\}$. The first four states are singly-occupied, while the last two are doubly occupied. The Fermi-Hubbard Hamiltonian above can be written as a matrix in this basis,
\begin{equation}
\hat{H}' = \left(
\begin{array}{cccccc}
 0 & 0 & 0 & 0 & 0 & 0 \\
 0 & 0 & 0 & 0 & -t_z e^{\frac{i \phi }{2}} & -t_z e^{\frac{i \phi }{2}} \\
 0 & 0 & 0 & 0 &  t_z e^{-\frac{i \phi }{2}} & t_z e^{-\frac{i \phi }{2}} \\
 0 & 0 & 0 & 0 & 0 & 0 \\
 0 & -t_z e^{-\frac{i \phi }{2}} & t_z e^{\frac{i \phi }{2}} & 0 & U - \eta_z [2(j-j_0) + 1] & 0 \\
 0 & -t_z e^{-\frac{i \phi }{2}} & t_z e^{\frac{i \phi }{2}} & 0 & 0 & U + \eta_z [2(j-j_0)+ 1] \\
\end{array}
\right).
\end{equation}
This Hamiltonian is broken into the the diagonal energies and the tunneling as a perturbation, $\hat{H}' = \hat{H}'_0 + \hat{H}'_{\mathrm{pert}}$ with,
\begin{equation}
\hat{H}'_0 = \left(
\begin{array}{cccccc}
 0 & 0 & 0 & 0 & 0 & 0 \\
 0 & 0 & 0 & 0 & 0 & 0\\
 0 & 0 & 0 & 0 &  0 & 0 \\
 0 & 0 & 0 & 0 & 0 & 0 \\
 0 & 0 & 0 & 0 & U - \eta_z [2(j-j_0) + 1] & 0 \\
 0 & 0 & 0 & 0 & 0 & U + \eta_z [2(j-j_0)+ 1] \\
\end{array}
\right),\>\>\>\>\hat{H}'_{\mathrm{pert}} = \left(
\begin{array}{cccccc}
 0 & 0 & 0 & 0 & 0 & 0 \\
 0 & 0 & 0 & 0 & -t_z e^{\frac{i \phi }{2}} & -t_z e^{\frac{i \phi }{2}} \\
 0 & 0 & 0 & 0 &  t_z e^{-\frac{i \phi }{2}} & t_z e^{-\frac{i \phi }{2}} \\
 0 & 0 & 0 & 0 & 0 & 0 \\
 0 & -t_z e^{-\frac{i \phi }{2}} & t_z e^{\frac{i \phi }{2}} & 0 & 0 & 0 \\
 0 & -t_z e^{-\frac{i \phi }{2}} & t_z e^{\frac{i \phi }{2}} & 0 & 0 & 0 \\
\end{array}
\right).
\end{equation}
The Schrieffer-Wolff generator is given by,
\begin{equation}
\hat{S} = \sum_{i,i'}\frac{\bra{i}\hat{H}'_{\mathrm{pert}}\ket{i'}}{\bra{i}\hat{H}'_0\ket{i} - \bra{i'}\hat{H}'_0\ket{i'}} \ket{i}\bra{i'},
\end{equation}
with $i$, $i'$ running over all 6 basis vectors of the matrix. The effective spin Hamiltonian is obtained from the generator via,
\begin{equation}
\hat{H}'_{\mathrm{Spin-1/2}} =\hat{P}_0 \hat{H}'_{0}\hat{P}_0 + \frac{1}{2} \hat{P}_0[\hat{S},\hat{H}'_{\mathrm{pert}}]\hat{P}_0,
\end{equation}
where $\hat{P}_0$ is the projector onto the singly occupied subspace,
\begin{equation}
\begin{aligned}
\hat{P} = \left(\begin{array}{cccccc}1 & 0 & 0 & 0 & 0 & 0 \\
0 & 1 & 0 & 0 & 0 & 0\\
0 & 0 & 1 & 0 & 0 & 0\\
0 & 0 & 0 & 1 & 0 & 0\\
0 & 0 & 0 & 0 & 0 & 0\\
0 & 0 & 0 & 0 & 0 & 0\end{array}\right).
\end{aligned}
\end{equation}
Keeping only the first four singly-occupied states in the resulting matrix, we obtain,
\begin{equation}
\begin{aligned}
\hat{H}'_{\mathrm{Spin-1/2}} = \frac{V_j}{4} \left(\begin{array}{cccc}1 & 0 & 0 & 0\\
0 & -1 & 2 e^{i \phi} & 0\\
0 & 2 e^{-i \phi} & -1 & 0\\
0 & 0 & 0 & 1\\\end{array}\right),
\end{aligned}
\end{equation}
where the matrix element is,
\begin{equation}
V_j = \frac{4t_z^2 U}{U^2 - [2(j-j_0)+1]^2 \eta_z^2}.
\end{equation}
This $4 \times 4$ Hamiltonian matrix can be expressed in terms of spin operators for the two sites (i.e. tensor products of $2 \times 2$ spin operator matrices),
\begin{equation}
\hat{H}'_{\mathrm{Spin-1/2}} = V_j\left[\frac{1}{2}\left(e^{i \phi}\hat{s}_{j}^{+}\hat{s}_{j+1}^{-}+h.c.\right)+\hat{s}_{j}^{z}\hat{s}_{j+1}^{z}\right]
\end{equation}
Finally, for chains longer than two sites we simply sum over all such nearest-neighbour couplings, thus adding a $\sum_{j=1}^{L-1}$ and yielding the model in the main text. If the Rabi drive is turned on, there will be an additional Hamiltonian term, but since the drive is site-local, it can be expressed by projecting a single-site Hamiltonian into the spin basis directly, yielding $\Omega\sum_{j}\hat{s}_j^{y}$. This direct projection is valid provided the Rabi drive is not too strong compared to the local potential differences, $\Omega \ll |U \pm \eta_z \left[2(j-j_0)+1\right]|$.

For the experimental conditions modeled, filling fraction is often less than unity. A spin-1/2 mapping is not suitable for this regime; however, we can instead employ a $t-J$ model, which still prohibits doubly occupied sites, but allows for the presence of holes. In this case we map each site to a spin-1, with spin states $\{\ket{1}_j, \ket{-1}_j\}$ corresponding to an $e,g$ atom respectively, while spin state $\ket{0}_j$ corresponds to an empty site.

We derive the spin-1 model in the same manner as above. We again write the full Fermi-Hubbard model as a matrix for two sites $j$, $j+1$. In this case, we must consider a larger Hilbert space. We use a basis of states $\{\ket{e,e}$, $\ket{e,0}$, $\ket{e,g}$, $\ket{0,e}$, $\ket{0,0}$, $\ket{0,g}$, $\ket{g,e}$, $\ket{g,0}$, $\ket{g,g}$, $\ket{eg,0}$, $\ket{0,eg}\}$, for which the Hamiltonian reads,
\footnotesize
\begin{equation}
\hat{H}' = 
\left(
\begin{array}{ccccccccccc}
 0 & 0 & 0 & 0 & 0 & 0 & 0 & 0 & 0 & 0 & 0 \\
 0 & -\eta_z\left(\delta j+1\right)^2 & 0 & -t_z e^{\frac{i \phi }{2}} & 0 & 0 & 0 & 0 & 0 & 0 & 0 \\
 0 & 0 & 0 & 0 & 0 & 0 & 0 & 0 & 0 & -t_z e^{\frac{i \phi }{2}} & -t_z e^{\frac{i \phi }{2}} \\
 0 & -t_z e^{-\frac{i \phi }{2}} & 0 & -\eta_z \delta j^2 & 0 & 0 & 0 & 0 & 0 & 0 & 0 \\
 0 & 0 & 0 & 0 & -\eta _z \delta j^2-\eta_z\left(\delta j+1\right)^2 & 0 & 0 & 0 & 0 & 0 & 0 \\
 0 & 0 & 0 & 0 & 0 & -\eta_z \delta j^2 & 0 & -t_z e^{\frac{i \phi }{2}}& 0 & 0 & 0 \\
 0 & 0 & 0 & 0 & 0 & 0 & 0 & 0 & 0 & t_z e^{-\frac{i \phi }{2}} & t_z e^{-\frac{i \phi }{2}} \\
 0 & 0 & 0 & 0 & 0 & -t_z e^{-\frac{i \phi }{2}} & 0 & -\eta_z \left(\delta j+1\right)^2 & 0 & 0 & 0 \\
 0 & 0 & 0 & 0 & 0 & 0 & 0 & 0 & 0 & 0 & 0 \\
 0 & 0 & -t_z e^{-\frac{i \phi }{2}} & 0 & 0 & 0 & t_z e^{\frac{i \phi }{2}} & 0 & 0 & U-\eta_z \left(2 \delta j+1\right) & 0 \\
 0 & 0 & -t_z e^{-\frac{i \phi }{2}} & 0 & 0 & 0 & t_z e^{\frac{i \phi }{2}} & 0 & 0 & 0 & U+ \eta_z\left(2 \delta j+1\right) \\
\end{array}
\right),
\end{equation}
\normalsize
where $\delta j=j-j_0$.

The first nine states $\{\ket{e,e}$,$ \ket{e,0}$,$\ket{e,g}$,$\ket{0,e}$,$\ket{0,0}$,$\ket{0,g}$,$\ket{g,e}$,$\ket{g,0}$,$\ket{g,g}\}$ of the Hilbert space comprise the spin-1 Hilbert space of non-doubly-occupied sites we are interested in, and are ordered in the same way as a natural tensor product structure of two spin-1 degrees of freedom. We take the top left $9 \times 9$ block and the bottom right $2 \times 2$ block of the matrix as the zeroth-order Hamiltonian $\hat{H}'_0$. All remaining off-diagonal tunneling matrix elements comprise the perturbation $\hat{H}'_{\mathrm{pert}}$. The projector $\hat{P}$ is onto the first 9 states. The $9 \times 9$ matrix resulting from the perturbation theory calculation is written in terms of tensor products of spin-1 operators (in practice this is done by using tensor products of orthogonal Gell-Mann matrices, then rewriting those in terms of canonical spin-1 operators). The resulting interaction is summed over all neighbouring pairs, which after some simplifying algebra yields our spin-1 $t_{z}-J$ model:
\begin{equation}
\begin{aligned}
\hat{H}'_{\mathrm{Spin-1}} &= -t_z \sum_{j}\left(e^{i \phi/2} \hat{S}_{j}^{+}\hat{S}_{j+1}^{-} +h.c.\right)\hat{S}_{j}^{z}\hat{S}_{j+1}^{z}+\frac{t_z}{2}\sum_{j}\left(e^{i \phi/2} \hat{S}_{j}^{+}\hat{S}_{j+1}^{-} - h.c.\right)\left(\hat{S}_{j}^{z} - \hat{S}_{j+1}^{z}\right)\\
&+\sum_{j}\frac{t_z^2 U}{U^2-[2(j-j_0)+1]^2 \eta_z^2}\big(\frac{e^{i \phi}}{2}\hat{S}_{j}^{+}\hat{S}_{j}^{+}\hat{S}_{j+1}^{-}\hat{S}_{j+1}^{-}+\frac{e^{-i \phi}}{2}\hat{S}_{j}^{-}\hat{S}_{j}^{-}\hat{S}_{j+1}^{+}\hat{S}_{j+1}^{+}\\
&\quad\quad\quad\quad\quad\quad\quad\quad\quad\quad\quad\quad\quad+\hat{S}_{j}^{z}\hat{S}_{j+1}^{z} - \hat{S}_{j}^{z}\hat{S}_{j}^{z} \hat{S}_{j+1}^{z}\hat{S}_{j+1}^z\big).
\end{aligned}
\end{equation}
This is the model used in the main text $t-J$ model simulations. If there is also a Rabi drive, its spin description is again obtained by projecting a single-site laser drive Hamiltonian into the spin-1 basis and summing over all $j$, yielding $\frac{-i \Omega}{4}\sum_j \left(\hat{S}_j^{+}\hat{S}_{j}^{+} - h.c.\right)$.

Our derivation omits the possibility of interaction-mediated double hops. Such processes are when an atom virtually hops into an already-occupied neighbouring site, then hops into another unoccupied site further along rather than returning to the site it started in - for instance going from $\ket{g,e,0}$ to $\ket{0,e,g}$ in a three-site chain. One can derive effective spin descriptions for such processes as well, writing the Hamiltonian matrix for three sites $j$, $j+1$, $j+2$ and proceeding as before. In our case, we have found that such additional terms do not contribute a significant correction.

%%%%%
\section{Lattice loading and filling fraction}
\label{app_Filling}
%%%%%

As emphasized throughout the paper the lattice density distribution, determining the local filling fractions, is a critical experimental parameter. In this section, we outline the lattice loading protocol and resulting filling fraction. A hyperfine spin-polarized atomic Fermi gas is initialized in a optical dipole trap (ODT), which generates simple harmonic confinement with trapping frequencies $\omega_{x}$, $\omega_y$, $\omega_z$; the lattice starts initially turned off. We use numbers from the experiment in Ref.~\cite{milner_science}: $\omega_x/(2\pi) = \omega_y/(2\pi) = 150$ Hz, $\omega_{z}/(2\pi) = 250$ Hz. After the cloud is cooled, the ODT is instead turned off, while the lattice is raised to the final depth used in studies of superexchange dynamics; a more detailed discussion of these ramps is given at the end of this Appendix. Throughout loading, the gas is assumed to be spin-polarized (there is no spin degree of freedom yet, all atoms are in their ground electronic spin state $g$).

The lattice filling fraction is primarily determined by the entropy-per-particle of the gas. We assume a Fermi-Dirac distribution in the lowest band of the lattice, with the energy set by the ODT trap,
\begin{equation}
\label{eq_FermiDirac}
P(x,y,z) = \frac{1}{e^{(E(x,y,z)-\mu)/(k_B T)}+1}
\end{equation}
where $x,y,z$ is the (integer) lattice position in units of the lattice spacing, $T$ is the gas temperature, $k_B$ is Boltzmann's constant, $\mu$ is a chemical potential offset, and $E(x,y,z)$ is the local potential energy from the ODT, parameterized in terms of the lattice constant $a$:
\begin{equation}
E(x,y,z) = \frac{1}{2}m \omega_{x}^2 a^2 x^2 + \frac{1}{2}m \omega_{y}^2 a^2 y^2 + \frac{1}{2}m \omega_{z}^2 a^2 z^2.
\end{equation}
While the ODT potential is continuous, we assume that raising the lattice forces all atoms to occupy only the potential minima at integer values of $x,y,z$. This approach employs the local density approximation, neglecting any tunneling effects or band excitations during the loading, and is a valid model for thermometry in the `atomic limit' for deep trap depths. The highest filling occurs in the center of the trap. For a spin polarized gas of fermions, the system is non-interacting $(U = 0)$ and the only relevant energy term is the confinement. We omit gravity along $z$ by assuming that it is incorporated into the definition of the $z$-direction potential; i.e. the minimum of the combined ODT + gravity is at $z = 0$, and the overall center of the potential is at $x=y=z=0$ during the loading process. The imaging techniques to retrieve this density distribution are detailed in Ref.~\cite{milner2023high, hutson2023observation}. 

The total number of atoms is given by the integral of the probability density,
\begin{equation}
\begin{aligned}
\label{eq_atomNumber}
N &= \int_{-\infty}^{\infty} dx\> \int_{-\infty}^{\infty} dy\> \int_{\infty}^{\infty} dz\> P(x,y,z) \\
&= - \left(\frac{2\pi k_B T}{m a^2}\right)^{3/2} \frac{1}{\omega_{x} \omega_{y} \omega_{z}} \text{Li}_{3/2} \left[-e^{\mu/(k_B T)}\right],
\end{aligned}
\end{equation}
where $\text{Li}_{n}[.]$ is the polylogarithm function. For fixed trapping frequencies and atom number $N$, the gas is described by just the temperature $T$; the chemical potential offset $\mu$ can obtained by numerically solving the above equation. Note that $\mu$ here is not the total chemical potential of the gas, but rather an offset relative to the depth at the bottom of the harmonic trap. In practice, the entropy of the gas will increase after ramping on the lattice due to non-adiabaticities in the loading process. While such effects could be modelled in theory, for our purposes we assume the temperature / entropy is empirically determined by experimental calibration. We emphasize that under the above model, the singular effect of temperature is to determine the density distribution in the lattice for atoms in the ground electronic state. The electronic spin degree of freedom is entirely decoupled from temperature effects. 

For spectroscopy, the ODT is ramped off and the atoms are confined in the 3D optical lattice operating at the magic wavelength. Since we work in the regime where the final $V_x,V_y$ lattice depths after ramping on the lattice are much higher than $V_z$, the subsequent dynamics of each vertical tube along $z$ can be solved separately, sampling initial conditions from a 1D Fermi-Dirac distribution rather than 3D. Eq.~\eqref{eq_FermiDirac} may still be used, with the $x$ and $y$ coordinates set to be constants depending on the $(x,y)$ position of the tube. 

To determine the distribution of uninterrupted chain lengths in main text Fig.~\ref{fig_ContrastChains}(c), we start with a fixed atom number $N = 30,000$. For a fixed temperature $T$, we obtain the chemical potential $\mu$ by numerically solving Eq.~\eqref{eq_atomNumber}. We insert $\mu$ and $T$ into the Fermi-Dirac distribution of Eq.~\eqref{eq_FermiDirac}, and set $x = y = 0$ to obtain the density distribution $P(z)$ for a tube at the horizontal center of the trap. We sample random particle configurations from this distribution, and count the number of uninterrupted chains $N_L$ of each length $L$. A distribution of particles $\bullet 00\bullet\bullet\bullet0\bullet\bullet000\bullet\bullet$ (atom $\bullet$ and hole $0$) would have $N_1 =1$, $N_2=2$, $N_3 = 1$. Averaging over many random particle configurations yields the full distribution. For a zero-temperature state $T = 0$, there is only one uninterrupted chain at the bottom of the potential, whose length is given by $L = \text{lim}_{T \to 0}\int_{-\infty}^{\infty}dz P(0,0,z)$. Note that Fig.~\ref{fig_ContrastChains}(c) reports temperature in terms of the Fermi temperature $T_F$; this is obtained from the Fermi energy $T_F = \epsilon_F/k_B$, which is in turn obtained from  $\epsilon_F = \text{lim}_{T \to 0} \mu$.

% We can also obtain the total entropy per particle of the gas,

% \begin{equation}
% \frac{S}{N k_{B}} = -\frac{\mu}{k_{\mathrm{B}}T} + \frac{5}{2} \frac{\text{Li}_{5/2}\left[-e^{\mu/(k_B T)}\right]}{\text{Li}_{3/2}\left[-e^{\mu/(k_B T)}\right]}.
% \end{equation}

We now comment on the specific ramping sequence used to turn off the ODT and turn on the lattice. In Ref.~\cite{milner_science}, we employ a tailored ramp sequence to achieve high filling fractions, largely motivated by the ground and first excited bands not being fully energetically separated until $2.24 \; E_{r}$ in a 3D optical lattice.  First, we adiabatically decompress the $1064$ nm ODT trap to reduce $T_{F}$. Next, we ramp the $813$ nm magic wavelength optical lattice to $2.5 \; E_{r}$ where there is a finite bandgap. Then, we strongly compress the $1064$ nm trap so the peak filling $n a^{3} \approx 1$. When highly degenerate, $n$ is set by the Fermi wavevector $\hbar k_{F} = \sqrt{2 m E_{F}} = \hbar (6 \pi^2 n)^{1/3}$. We finally ramp the optical lattice to the final setpoint, where $t_{x}=t_y \ll 1$ Hz, effectively freezing the atom's density distribution.

%%%%%
\section{Confining potential}
\label{app_Potential}
%%%%%

In this Appendix, we detail the confining potentials and the associated cloud position, for both the ODT trap (used only during loading) and the confining potential induced by the lattice beam curvature (active during the superexchange induced dynamics for the Ramsey/Rabi protocols).

During loading, the atoms feel a vertical potential from gravity plus the ODT given by $V(z) = mgaz + \frac{1}{2}m \omega_z^2 a^2 z^2$, with gravity $g=9.81 m/s^2$, which has a minimum at $z_{\mathrm{ODT}}=-\frac{g}{a\omega_z^2}$. In our case, for $a = 406.5$ nm this is $z_{\mathrm{ODT}} \approx 9.8$ lattice sites, or roughly $z_{\mathrm{ODT}} a \approx 4 \mu m$ below the center of the ODT beam alone. The initial atomic cloud is vertically centered at this position.

After cooling the gas in the ODT, the lattice beams are ramped on. These beams also contribute a confining potential in addition to enforcing the discretized lattice spacing, with a different harmonic trapping frequency $\omega_{\mathrm{lat},z}$ (different from the ODT); the full derivation for this potential is provided at the end of this Appendix. For the experiment in Ref.~\cite{milner_science}, the lattice beams are aimed at the atomic cloud's center position \text{after} loading and cooling in the ODT, including the displacement due to gravity. Hence the lattice beams' harmonic potential is $V_{\mathrm{lat}}(z) = \frac{1}{2}m \omega_{\mathrm{lat,z}}^2a^2(z - z_{\mathrm{ODT}})^2$, with a vertical spatial offset by $z_{\mathrm{ODT}} \approx 9.8$ from the ODT beam. While both the ODT and lattice beams are on, the atoms at the center of the trap thus do not feel any additional potential curvature.

After the ODT is ramped off, the potential contains only the lattice induced harmonic confinement, and instead becomes $V(z) = mgaz + \frac{1}{2}m \omega_{\mathrm{lat},z}^2 a^2 \left(z - z_{\mathrm{ODT}}\right)^2$. The bottom of this new potential is now at $z = z_{\mathrm{ODT}}- \frac{g}{a\omega_{\mathrm{lat},z}^2}$. When compared to the original center of the cloud at $z_{\mathrm{ODT}}$, we find that the new harmonic trap center is below the cloud by a distance of
\begin{equation}
z_{\mathrm{\mathrm{lat}}} = -\frac{g}{a \omega_{\mathrm{lat,z}}^2} = -\frac{m g a}{2 \eta_z}
\end{equation}
lattice sites. For any uninterrupted chain of sites modeled by superexchange, the center of the lattice potential $j_0$ is thus shifted by the above value, in addition to wherever the chain itself was found according to the Fermi-Dirac distribution. In practice, this means that for a shallow lattice, the cloud will be shifted very far from the bottom of the trap. Note also that for this setup, the local potential \textit{difference} between adjacent sites at the center of the cloud is $\approx 2\eta_z z_{\mathrm{lat}} = -m g a$, i.e. the local potential curvature exactly cancels the gravitational sag. However, the cloud can also be moved around to other positions relative to the bottom of the lattice harmonic trap as discussed further in Ref.~\cite{milner_science}.

For clarity, here we also spell out the full derivation of the lattice potential, starting from a non-harmonic treatment. Unlike the ODT the lattice confining potential is present throughout the quantum dynamics, and beyond-harmonic effects can become relevant (though we do not thoroughly explore such physics in this work). The full 3D potential for the lattice reads,
\begin{equation}
V_{\mathrm{lat}}(x,y,z) = -V_x\cos^2\left(\pi x\right) e^{- 2a^2 \frac{y^2+z^2}{W_x^2}} -V_y \cos^2\left(\pi y\right) e^{- 2a^2 \frac{x^2+z^2}{W_y^2}} - V_{z} \cos^2\left(\pi z\right)e^{- 2a^2 \frac{x^2+y^2}{W_z^2}},
\end{equation}
with $V_x, V_y, V_z$ the lattice depths and $W_x, W_y, W_z$ the respective $x, y, z$ beam waists. The sinusoidal terms are the lattice potential due to the standing wave, while the Gaussian terms capture the curvature of the beams. Note that this is an attractive potential, so the lattice site wells are at integer values of $x,y,z$. Near the bottom of the trap ($a|x| \ll W_x, a|y| \ll W_y, a|z|\ll W_z$) the Gaussian terms can be Taylor expanded to give (dropping constant terms),
\begin{equation}
\begin{aligned}
V_{\mathrm{lat}}(x,y,z) &\approx - V_x \cos^2\left(\pi x\right) - V_y \cos^2\left(\pi y\right) - V_z \cos^2\left(\pi z\right)\\
&+\frac{2V_x a^2}{W_x^2}\left(y^2+z^2\right)\cos^2\left(\pi x\right) +\frac{2V_y a^2}{W_y^2}\left(x^2+z^2\right) \cos^2\left(\pi y\right) +\frac{2V_{z}}{W_z^2}\left(x^2+y^2\right) \cos^2\left(\pi z\right).
\end{aligned}
\end{equation}
At the horizontal center of the trap $x = y=0$, the potential simplifies to (dropping constant terms),
\begin{equation}
\begin{aligned}
V_{\mathrm{lat}}(0,0,z) &= - V_z \cos^2\left(\pi z\right)+\eta_z z^2,\\
\eta_z &=\left(\frac{2V_x a^2}{W_x^2}+\frac{2V_y a^2}{W_y^2}\right).
\end{aligned}
\end{equation}
The second term is precisely our harmonic potential along $z$, with $\eta_z$ the trap energy described in the main text. The harmonic trapping frequency can be obtained by writing,
\begin{equation}
\eta_z = \frac{1}{2}m \omega_{\mathrm{lat,z}}^2 a^2.
\end{equation}

%%%%%
\section{Analytic results for Ramsey and Rabi decay}
\label{app_Analytics}
%%%%%

While writing both the Ramsey and Rabi dynamics for generic chain length $L$ is analytically intractable, it can be done for small $L=2$ and $L=3$ chains with homogeneous couplings $V_j = V$ (only relevant for the $L=3$ chain which has two couplings). For Ramsey contrast decay we have,
\begin{equation}
\begin{aligned}
C_{L=1}(t) &= 1,\\
C_{L=2}(t) &= \left|\cos^2\left(\frac{\phi}{2}\right)+ \sin^2\left(\frac{\phi}{2}\right)\cos\left(V t\right)\right|,\\
C_{L=3}(t) &= \frac{1}{36}\bigg|26+15 \cos(\phi)+12 \cos(2\phi)+\cos(3\phi)+12[2+\cos(\phi)]\sin^2(\phi)\cos\left(\frac{V t}{2}\right)\\
&\quad\quad\>\>+24 \sin^2\left(\frac{\phi}{2}\right)\sin^2(\phi)\cos(V t)+32 \sin^6\left(\frac{\phi}{2}\right)\cos\left(\frac{3V t}{2}\right)\bigg|.
\end{aligned}
\end{equation}
Notice that for $L=3$ there are three different oscillatory frequencies $\sim V/2$, $V$, $3V/2$, which stem from contributions of different total angular momentum sectors of the Hilbert space.

For Rabi dynamics, unfortunately even the $L=2$ chain is too intractable to write the full solution analytically. Series expansions in small $\Omega/V$ also fail to capture the longer-time behavior (namely the sporadic oscillations or lack thereof) observed numerically. However, for modest sized $L$ it is still straightforward to solve the quantum dynamics numerically.

%%%%%
\section{Mean-field theory}
\label{app_MeanField}
%%%%%

Below are the mean-field Heisenberg equations of motion for one-point spin-1/2 observables under the superexchange Hamiltonian with a non-zero Rabi drive, assuming homogeneous couplings $V_j = V$.
\begin{equation}
\begin{aligned}
\frac{d}{dt}\langle\hat{s}_j^{+}\rangle &= - i V\left(e^{i \phi} \langle \hat{s}_{j-1}^{+}\rangle + e^{-i \phi}\langle \hat{s}_{j+1}^{+}\rangle \right) \langle\hat{s}_j^{z}\rangle + i V \left(\langle\hat{s}_{j-1}^{z}\rangle + \langle\hat{s}_{j+1}^{z}\rangle\right)\langle\hat{s}_{j}^{+}\rangle + \Omega \langle\hat{s}_{j}^{z}\rangle,\\
\frac{d}{dt}\langle\hat{s}_j^{-}\rangle &=  +i V\left(e^{-i \phi} \langle \hat{s}_{j-1}^{-}\rangle + e^{i \phi}\langle \hat{s}_{j+1}^{-}\rangle \right) \langle\hat{s}_j^{z}\rangle - i V \left(\langle\hat{s}_{j-1}^{z}\rangle + \langle\hat{s}_{j+1}^{z}\rangle\right)\langle\hat{s}_{j}^{-}\rangle + \Omega \langle\hat{s}_{j}^{z}\rangle,\\
\frac{d}{dt}\langle\hat{s}_j^{z}\rangle &= -i \frac{V}{2} \left(e^{-i \phi}\langle \hat{s}_{j-1}^{-}\rangle + e^{i \phi}\langle \hat{s}_{j+1}^{-}\rangle\right)\langle\hat{s}_{j}^{+}\rangle + i \frac{V}{2} \left(e^{i \phi}\langle \hat{s}_{j-1}^{+}\rangle + e^{-i \phi}\langle \hat{s}_{j+1}^{+}\rangle\right)\langle\hat{s}_{j}^{-}\rangle - \frac{\Omega}{2}\left(\langle\hat{s}_j^{+} \rangle + \langle\hat{s}_j^{-} \rangle\right).
\end{aligned}
\end{equation}
These equations are straightforward to solve numerically, as they scale linearly in $L$.

\end{document}